\documentclass[fleqn,usenatbib]{mnras}

\usepackage{newtxtext,newtxmath,hyperref}
\usepackage[T1]{fontenc}
\usepackage{graphicx}	
\usepackage{amsmath}	
\usepackage{amssymb}	
\usepackage{enumitem}
\usepackage[english]{babel}
\usepackage{gensymb}

\title[Redundant-Baseline Calibration Without Adding Spectral Structure]{Polarized Redundant-Baseline Calibration for 21\,cm Cosmology Without Adding Spectral Structure}

\author[J.~S. Dillon et al.]{Joshua S. Dillon,$^{1\dagger}$\thanks{E-mail: jsdillon@berkeley.edu}
Saul A. Kohn,$^{2}$
Aaron R. Parsons,$^{1}$
James E. Aguirre,$^{2}$ \newauthor
Zaki S. Ali,$^{1}$ 
Gianni Bernardi,$^{3,4,5}$ 
Nicholas S. Kern,$^{1}$
Wenyang Li,$^{6}$ 
Adrian Liu,$^{1\ddagger}$ \newauthor
Chuneeta D. Nunhokee,$^{1,4}$
Jonathan C. Pober$^{6}$
\\
$^{1}$Department of Astronomy, University of California, Berkeley, Berkeley, CA, USA\\
$^{2}$Department of Physics and Astronomy, University of Pennsylvania, Philadelphia, PA, USA\\
$^{3}$INAF-Istituto di Radioastronomia, Bologna, Italy\\
$^{4}$Department of Physics and Electronics, Rhodes University, Grahamstown, South Africa\\
$^{5}$Square Kilometre Array South Africa, Cape Town, South Africa \\
$^{6}$Department of Physics, Brown University, Providence, RI, USA\\
$^{\dagger}$NSF Astronomy and Astrophysics Postdoctoral Fellow\\
$^{\ddagger}$Hubble Fellow
}

\date{\today}
\pubyear{2017}

\begin{document}
\label{firstpage}
\pagerange{\pageref{firstpage}--\pageref{lastpage}}
\maketitle

\begin{abstract}
21\,cm cosmology is a promising new probe of the evolution of visible matter in our universe, especially during the poorly-constrained Cosmic Dawn and Epoch of Reionization. However, in order to separate the 21\,cm signal from bright astrophysical foregrounds, we need an exquisite understanding of our telescopes so as to avoid adding spectral structure to spectrally-smooth foregrounds. One powerful calibration method relies on repeated simultaneous measurements of the same interferometric baseline to solve for the sky signal and for instrumental parameters simultaneously. However, certain degrees of freedom are not constrained by asserting internal consistency between redundant measurements. In this paper, we review the origin of these \emph{degeneracies} of redundant-baseline calibration and demonstrate how they can source unwanted spectral structure in our measurement and show how to eliminate that additional, artificial structure. We also generalize redundant calibration to dual-polarization instruments, derive the degeneracy structure, and explore the unique challenges to calibration and preserving spectral smoothness presented by a polarized measurement.
\end{abstract}



\section{Introduction} \label{sec:intro}

21\,cm cosmology---the mapping of neutral hydrogen in the universe using redshifted 21\,cm hyperfine emission or absorption---promises unprecedented insight into the intergalactic medium and its density, ionization, thermal, and velocity structure. By probing that evolution, we can better understand the astrophysics of the first stars, galaxies, and black holes that eventually led to the Epoch of Reionization (EoR) \citep{FurlanettoReview, miguelreview, PritchardLoebReview, SaleemEoRChapter, aviBook} and precisely test our cosmological theories at both low- \citep{ChangDE,wyithe2008} and high-redshifts \citep{Matt3,Yi}.

Realizing that promise requires overcoming two key challenges. The first is that 21\,cm fluctuations are predicted to be extremely faint, requiring massive radio telescopes to detect. \citet{GBT} used the Green Bank Telescope to detect the signal in cross-correlation with a galaxy survey. The pioneering generation of EoR observatories---including the Low Frequency Array (LOFAR; \citealt{LOFARLimitsPatil}), the Giant Metrewave Radio Telescope (GMRT; \citealt{newGMRT}), the Murchison Widefield Array (MWA; \citealt{EmpiricalCovariance,CHIPS,BeardsleyFirstSeason}), and the Donald C. Backer Precision Array for Probing the Epoch of Reionization (PAPER; \citealt{PAPER64Limits})---have put upper limits on the signal, but it is very difficult for them to achieve more than a tentative detection. That is why the next generation telescopes have invested in vastly increased collecting area, including the Canadian Hydrogen Intensity Mapping Experiment (CHIME; \citealt{CHIMEpathfinder}) and the Hydrogen Intensity and Real-time Analysis eXperiment (HIRAX; \citealt{HIRAXconcept}) at low $z$ and the Hydrogen Epoch of Reionization Array (HERA; \citealt{RedArrayConfig,HERAOverview}) and the Square Kilometer Array (SKA-low; \citealt{LeonCosmicDawnEoRSKA}) at high $z$.

The second and perhaps greater challenge is avoiding the contamination of the cosmological signal by astrophysical foregrounds ${\sim}10^5$ times brighter \citep{Santos, Jelic08, BernardiForegrounds, GhoshForegrounds, SaulPAPERPol}. While foregrounds are spectrally very smooth, the 21\,cm brightness temperate fluctuations are expected to exhibit structure on fine $\Delta z$ (and thus fine frequency) scales. The strategy of excising or down-weighting foregrounds by simply excising a few low-$k_\|$ Fourier modes \citep{LT11,ChapmanGMCA,DillonFast,X13,BonaldiCCA} is problematized by the spectrally complex response of any interferometer, which generically features a position- and frequency-dependent point-spread function that is very difficult to invert accurately \citep{mapmaking}. 

Fortunately, this effect on typical power-law foregrounds is limited to a region of 2-D Fourier space (i.e. $k_\|$ and $k_\perp$) known as the \emph{wedge} \citep{Dattapowerspec,AaronDelay,VedanthamWedge,MoralesPSShapes,Hazelton2013,PoberWedge,ThyagarajanWedge,EoRWindow1,EoRWindow2}. Avoiding foregrounds by working outside the wedge (in the so-called \emph{EoR window}) can achieve more robust foreground isolation, but at the cost of losing considerable sensitivity \citep{PoberNextGen}. This separation works as long as foregrounds are intrinsically spectrally smooth and the instrument does not impart additional spectral structure.

The instrumental response to polarized foregrounds is another potential cause for concern. 21\,cm experiments largely concentrate on unpolarized (Stokes~$I$) measurements because the cosmological signal is essentially unpolarized \citep{21cmGravWaveCircularForecast}. 
However, all instruments have frequency- and direction-dependent responses to linear polarization (Stokes~$Q$ and Stokes~$U$) and circular polarization (Stokes~$V$) that can be difficult to disentangle from unpolarized emission without an exquisite instrument model \citep{Jones1941}. Galactic Faraday rotation can turn foreground Stokes~$Q$ into Stokes~$U$ and back as a function of frequency---often many times over the band of interest. Since Stokes~$Q$ and Stokes~$U$ generally leak differently into our estimate of Stokes~$I$, this can introduce spectral structure into the observed foregrounds that depends on the rotation measure. A $Q$- or $U$-contaminated estimate of Stokes~$I$ power spectrum will show leakage from low $k_\|$ (inside the wedge) to higher $k_\|$, potentially introducing a bias to measurements made only inside the EoR window \citep{JelicRealistic,MoorePolarization,AsadPolarization1,EmilPol,NunhokeePolarized}. Measurements made to quantify the magnitude of the effect (e.g.\ \citealt{SaulPAPERPol,MoorePolLimits,AsadPolarization3}) generally show that, to current sensitivity levels, it is not a show-stopper for EoR science. 

Fundamentally, any solution to the problem of foregrounds---polarized or unpolarized---relies on accurate instrumental knowledge. Knowing antenna locations and their polarized, frequency-dependent primary beam response functions is key to both foreground subtraction and to making Stokes-$I$ measurements free from polarization leakage \citep{Richard, ShawCoaxing}.

Accurate instrumental knowledge requires an accurate calibration of our analog signal chains. Complex antenna bandpasses with complex spectral structure, if not accurately measured and accounted for, can scatter foreground power outside the wedge far in excess of the cosmological signal. As an example: an uncalibrated ${\sim}1\%$ sinusoidal ripple in visibilities due to cable reflections can produce foreground contamination $\sim$$10^6$ times larger than the nominal power spectrum signal, obliterating the EoR window \citep{AaronFirstEoXLimits}.

Traditionally, antenna bandpass calibration has depended upon producing an accurate sky model through iterative cycles of mapmaking, source extraction, and re-calibration \citep{selfcal,InitialLOFAR1}. However, any realistic level of error in one's sky or instrument model leads to gain calibration errors. These errors are chromatic because the visibilities used to create them have spectral structure that depends on their length (among other factors). Since the same antennas are involved in both long and short baselines, short baseline measurements inherit chromatic errors from long baselines with more intrinsic spectral structure. This leakage of spectral structure via the gain errors can dramatically restrict the size of the EoR window and thus the sensitivity to the 21\,cm signal. Avoiding this effect requires either assumptions about the spectral and temporal smoothness of antenna bandpasses \citep{YatawattaConsensus, BarryCal} or a reweighting of baselines \citep{ModelingNoise}. 

More recently, another approach has been employed to considerable success. Instead of calibrating by reference to a sky-model, arrays with highly-redundant configurations can use internal consistency between repeated measurements to solve for most of the calibration degrees of freedom. The idea was developed in the present formalism by \citet{redundant}, though it has its antecedents in older work (e.g.\ \citealt{Wieringa}). This \emph{redundant-baseline calibration} relies on a simple counting argument. Since baselines with the same separation between antennas measure an identical integral over the sky, if the number of instantaneous visibility measurements is significantly larger than the number of unique baselines, then it may be possible to solve for antenna-based calibration parameters and unique visibilities simultaneously. 

A number of 21\,cm arrays have been designed or were reconfigured to take advantage of redundant baseline calibration. This is in part due to the synergy with the many short baselines measuring the same modes over and over, which helps more economically meet the extraordinary sensitivity requirements for 21\,cm cosmology \citep{AaronSensitivity}. The technology demonstrator MITEoR pioneered the approach for EoR science, showing for the first time calibration residuals largely consistent with thermal noise  \citep{MITEoR}. PAPER was reconfigured in a redundant configuration for its 32-element observing season \citep{PAPER32Limits}. The MWA was recently expanded to include a pair of 36-element redundant hexes (\citeauthor{WenyangMWAHex}~\textit{in~review}). HERA was designed to be completely redundantly calibratable, including its outrigger antennas \citep{RedArrayConfig,HERAOverview}.

That said, redundant-baseline calibration is not the end of the story. Like any calibration scheme based on internal consistency, it falls a bit short of a complete solution for every signal-chain calibration parameter.\footnote{These are similar but not identical to the integer multiple of $\pi$ phase ambiguities that arise in self-calibration \citep{YatawattaAmbiguity,IntegerAmbiguity}.}  Roughly speaking, redundant-baseline calibration reduces the calibration problem from one number per frequency channel per antenna element to just a few numbers per frequency for the whole array. Still, the power of redundant calibration is that it vastly reduces the number of degrees of freedom in the calibration problem, yielding precise relative calibration and reducing the complexity of any future \emph{absolute calibration} referenced to the sky. 

The linear combinations of gains and visibilities that redundant-baseline calibration cannot solve for, the \emph{degeneracies} of its underlying system of equations, have been the source of much confusion. What exactly can redundant-baseline calibration solve for? And what are the degeneracies? In this paper, we present a pedagogical explanation of the degeneracies and show their importance by demonstrating the pitfalls they may present. For redundant-baseline calibration to be useful for 21\,cm cosmology, it must preserve the spectral smoothness of the observation and it must be possible to understand and undo any modifications made to the degenerate part of the calibration solutions. In principle, any two solutions that differ only in their degeneracies can be transformed and be brought in line with a good absolute calibration by reference to the sky. In practice, this can be somewhat perilous---especially if one wants to use prior knowledge about the instrument to restrict the spectral degrees of freedom in bandpass calibration in a way that enforces relative smoothness and cannot account for discontinuities in frequency. Redundant-baseline calibration as a single modular analysis step has seen increasing use recently (e.g.\ \citealt{MITEoR,ZhengBruteForce,PAPER64Limits}) and therefore it is valuable to consider the subtleties of maintaining spectral smoothness.

In this paper we show how the degeneracies be cleanly and self-consistently accounted for so that the resultant calibration solutions can later be combined with a sky-referenced absolute calibration. In Section~\ref{sec:pedagogical} we review the mathematical formalism underpinning redundant-baseline calibration, explain the mathematical origin of the degeneracies, and show how careless handling of the degeneracies can lead to considerable spectral structure. 
Then, in Section~\ref{sec:polarization}, we generalize the derivation of degeneracies to observations with dual-polarized antennas and highlight new complications and analysis choices that arise.


\section{A Pedagogical Review of Redundant-Baseline Calibration} \label{sec:pedagogical}

We begin this section with a review of the antenna calibration problem in the context of a single visibility polarization and how it can be addressed by taking advantage of redundant baselines (Section~\ref{sec:calproblem}). Drawing heavily on \citet{redundant}, we then present an iterative algorithm for minimizing the error in our solution (Section~\ref{sec:linearizing}). Next we explain what exactly redundant-baseline calibration can and cannot solve for (Section~\ref{sec:1poldegen}) and how these unsolvable quantities can introduce spectral structure into our calibration solutions if we are not careful (Section~\ref{sec:delay_errors}). Finally we step back with an overview of the key assumptions that underlie the use of redundant-baseline calibration (Section~\ref{sec:assumptions}).


\subsection{The Calibration Problem}\label{sec:calproblem}

Fundamentally, the problem of calibration\footnote{For the purposes of this work, we mean signal-chain calibration of antenna-based gains. This is sometimes called \emph{direction-independent} calibration to contrast it with \emph{direction-dependent} calibration that corrects for the shape and possible antenna-to-antenna variation of the primary beam response. The two effects can always be factored, though it can be useful to consider them together in the context of sky-based calibration.} boils down to one key equation:
\begin{equation}
V_{ij}^\text{obs}(\nu) = g_i(\nu) g_j^*(\nu) V_{ij}^\text{true}(\nu) + n_{ij}(\nu).
\end{equation}
The observed visibility $V_{ij}$ between antennas $i$ and $j$ at a given time and frequency is related to the true, underlying visibility by a pair of complex and frequency-dependent gain factors, $g_i$ and $g_j^*$, along with Gaussian random noise $n_{ij}$. These gain factors incorporate the frequency response of the analog signal chain, including amplifiers, attenuators, cable reflections, and of course the phase factor $e^{-2\pi i\tau\nu}$ due to the light-travel-time delay $\tau$ along the signal path. When we calibrate, we want to solve for both those gains and visibilities. In essence, we have a system of equations for all antenna pairs $i$ and $j$ given by
\begin{equation} \label{eq:system}
V_{ij}^\text{obs}(\nu) = g_i(\nu) g_j^*(\nu) V_{i-j}^\text{sol}(\nu)
\end{equation}
and we want to find the optimal gain and visibility solutions that minimize the error in this system.

In equation~\ref{eq:system}, we write our visibility $V_{i-j}$ as shorthand for $V(\mathbf{r}_i - \mathbf{r}_j)$, the visibility for the baseline vector $\mathbf{b}_{ij} \equiv \mathbf{r}_i - \mathbf{r}_j$. Two pairs of identical elements with precisely the same baseline separation between are sensitive to the same mode on the sky. The measured visibilities may be quite different due to differences in signal chains between the four elements, but that fact is incorporated into the gains. In a highly-redundant array, there are many more measurements than there are unique baselines.

Consider the example of HERA, which is made of 14\,m dishes with a core of 320 hexagonally-packed elements. The simplest proposed configuration of a HERA-like instrument---a densely packed hexagonal array of 331 elements (11 on each edge)---has $331(331-1)/2 = 54,615$ baselines and thus measures 54,615 visibilities. However, it only has 630 unique baseline separations \citep{RedArrayConfig}. That means we have a non-linear system of 54,615 equations to determine 630 unique visibilities and the 331 complex gains; the system is vastly overdetermined.\footnote{Ignoring for the moment the degeneracies in this system of equations that we will return to in Section~\ref{sec:1poldegen}.} With uncorrelated Gaussian noise and an overdetermined system of equations, it is useful to minimize $\chi^2$, defined as
\begin{equation} \label{eq:chisq}
\chi^2(\nu) = \sum_{\text{all pairs }i,j} \frac{\left|V_{ij}^\text{obs}(\nu) - g_i(\nu) g_j^*(\nu) V^\text{sol}_{i-j}(\nu)\right|^2}{\sigma^2_{ij}(\nu)}
\end{equation}
where $\sigma^2_{ij}(\nu)$ is the variance of $n_{ij}(\nu)$.  

If all one wishes to minimize is the difference between the calibration solution and the data, equation~\ref{eq:chisq} is sufficient. However, one may also wish to impose additional prior information about the sky or the instrument. To enforce that gains are relatively spectrally smooth, as in the consensus optimization of \cite{YatawattaConsensus}, one could add a penalty factor $\chi^2$ for gain discontinuities between nearby frequency channels.\footnote{Alternatively, one can take the approach of \citet{MITEoR} and use a Weiner filter to smooth the gains after redundant-baseline calibration based on an estimate of how much time-to-time and frequency-to-frequency variation is due to thermal noise. This will generally not produce the same result the \citet{YatawattaConsensus} method, since it smooths about a different $\chi^2$ minimum---one without that penalty factor for non-smoothness in the gains.} Likewise, \citet{CorrCal} proposes adding information about sky sources and their statistics into $\chi^2$, as well as information about known deviations from redundancy between putatively identical baselines, to bridge the gap between sky-based calibration and redundant calibration. That said, our focus is the simplest and most straightforward approach to redundant-baseline calibration in which we aim to minimize the $\chi^2$ in equation~\ref{eq:chisq}. This non-linear least-squares optimization can be done independently between frequencies (and times, for that matter), so we drop the explicit dependence on $\nu$ for now.


\subsection{Linearizing and Minimizing $\chi^2$} \label{sec:linearizing}

There are many methods for finding set of gains and visibilities that minimize $\chi^2$. They generally involve linearizing equation~\ref{eq:system} in order to find parameter solutions, often iteratively. \citet{redundant} suggests two. The first involves linearizing equation~\ref{eq:system} by taking the logarithm of both sides and then solving independently for the real and imaginary parts of the gains and visibilities. Unfortunately, this method produces biased solutions that do not actually minimize $\chi^2$.

That is why \citet{redundant} also advance a linearized method that implements the Gauss-Newton algorithm. Given some starting gains $g^0_i$ and visibilities $V^0_{i-j}$, we can rewrite equation~\ref{eq:system} as
\begin{equation}
V_{ij}^\text{obs} = (g^0_i + \Delta g_i) (g^{0}_j + \Delta g_j)^* (V^0_{i-j} + \Delta V_{i-j}).
\end{equation}
Here, the $\Delta$ terms are solved for at each iteration, allowing us to update our guesses until we converge on the global minimum. If we assume they are all small because our initial guess is close, we drop the $\Delta^2$ terms and get a linear system of equations for the $\Delta$ terms:
\begin{equation} \label{eq:linearized}
V_{ij}^\text{obs} - g^0_i g^{0*}_j V^0_{i-j} = \Delta g_i g^{0*}_j V^0_{i-j} + g^{0}_i \Delta g_j^*  V^0_{i-j} +  g^0_i g^{0*}_j \Delta V_{i-j}.
\end{equation}

The complex conjugation of the system in equation~\ref{eq:linearized} requires us to break it into real and imaginary parts in order to write the system as a matrix. This yields
\newcommand{\Real}[1]{\mathbb{R}\text{e}\hspace{-1pt}\left[#1\right]}
\newcommand{\Imag}[1]{\mathbb{I}\text{m}\hspace{-1pt}\left[#1\right]}
\begin{align} \label{eq:real}
\mathbb{R}\text{e} \left[ V_{ij}^\text{obs} - \right. & \left. g^0_i g^{0*}_j V^0_{i-j} \right] = \nonumber \\
&\Real{\Delta g_i} \Real{g^{0*}_j V^0_{i-j}} - \Imag{\Delta g_i} \Imag{g^{0*}_j V^0_{i-j}} + \nonumber \\
&\Real{\Delta g_j} \Real{g^{0}_i V^0_{i-j}} + \Imag{\Delta g_j} \Imag{g^{0}_i V^0_{i-j}} + \nonumber \\
&\Real{\Delta V_{i-j}} \Real{g^0_i g^{0*}_j} - \Imag{\Delta V_{i-j}} \Imag{g^0_i g^{0*}_j} 
\end{align}
and
\begin{align} \label{eq:imag}
\mathbb{I}\text{m} \left[ V_{ij}^\text{obs} - \right. & \left. g^0_i g^{0*}_j V^0_{i-j} \right] = \nonumber \\
&\Imag{\Delta g_i} \Real{g^{0*}_j V^0_{i-j}} + \Real{\Delta g_i} \Imag{g^{0*}_j V^0_{i-j}} + \nonumber \\
-&\Imag{\Delta g_j} \Real{g^{0}_i V^0_{i-j}} + \Real{\Delta g_j} \Imag{g^{0}_i V^0_{i-j}} + \nonumber \\
&\Imag{\Delta V_{i-j}} \Real{g^0_i g^{0*}_j} + \Real{\Delta V_{i-j}} \Imag{g^0_i g^{0*}_j}.
\end{align}
We can now write this set of equations compactly as
\begin{equation}
\mathbf{d} = \mathbf{A x}.
\end{equation}
Here $\mathbf{d}$ contains both the real and imaginary components of the differences between our guesses and $V_{ij}^\text{obs}$. $\mathbf{x}$ contains the real and imaginary parts of our $\Delta g_i$ and $\Delta V_{i-j}$ terms, and $\mathbf{A}$ contains all the coefficients in equations~\ref{eq:real} and \ref{eq:imag}. $\mathbf{d}$ has a length equal to twice the number of observed visibilities while $\mathbf{x}$ has a length equal to twice the sum of the number of unique visibilities and the number of antennas.

If $\mathbf{N}$ is noise covariance between visibility measurements, then the optimal estimate of the $\Delta$ terms is given by
\begin{equation} \label{eq:noiseInvEstimator}
\widehat{\mathbf{x}} = \left(\mathbf{A}^\intercal \mathbf{N}^{-1} \mathbf{A} \right)^{-1} \mathbf{A}^\intercal \mathbf{N}^{-1} \mathbf{d}.
\end{equation} 
Since visibility measurements have uncorrelated noise, $\mathbf{N}$ is diagonal and has the form $N_{ij,kl} = \sigma^2_{ij} \delta_{ik}\delta_{jl}$ where $\delta_{ik}$ is the Kronecker delta. However, if all baseline variances $\sigma^2_{ij}$ are identical, equation~\ref{eq:noiseInvEstimator} reduces to
\begin{equation} \label{eq:linearEstimator}
\widehat{\mathbf{x}} = \left(\mathbf{A}^\intercal \mathbf{A} \right)^{-1} \mathbf{A}^\intercal \mathbf{d}.
\end{equation} 
Thus, to find the set of gains and visibilities that minimizes $\chi^2$, we iteratively set up and solve equation~\ref{eq:linearEstimator} until we reach the desired level of convergence. In practice, this is complicated by the fact that 
$\mathbf{A}^\intercal \mathbf{A}$ is not invertible---a consequence of the inherent degeneracies in the system of equations that we will discuss at length in Section~\ref{sec:1poldegen}---and therefore requires a modified inversion technique like the  Moore-Penrose pseudoinverse.

This method is sufficient for understanding how the structure of $\mathbf{A}^\intercal \mathbf{A}$ relates to $\chi^2$, but we also want to emphasize that the Gauss-Newton method of \citet{redundant} is not the only approach to redundant-baseline calibration. \citet{LOFARcal} advance the weighted alternating least-squares technique, which linearizes equation~\ref{eq:system} by alternatingly holding gains and visibilities constant and zeroing-in on the minimum. Though this can take more steps to converge, it can also be much faster because it can cut the size of the matrix inversions, the rate limiting step in most least-squares minimizations. The \texttt{omnical} code developed for MITEoR \citep{MITEoR,ZhengBruteForce} is based on this approach.\footnote{The most recent version of \href{https://github.com/HERA-Team/hera_cal/tree/v1.0}{\texttt{hera\_cal}} has interfaces to both \href{https://github.com/HERA-Team/omnical/tree/cdf7e2f1f4249eac346ad13a47ca19a7a7229240}{\texttt{omnical}}, which is based on the alternating least-squares method, and \texttt{redcal}, which implements the full matrix-inversions of the Gauss-Newton method using the \href{https://github.com/HERA-Team/linsolve/tree/20b9958430e8fc9ab852deea383625832e1e3517}{\texttt{linsolve}} package. The former is much faster, but more vulnerable to false-minima (\citeauthor{WenyangMWAHex}~\textit{in~review}).} \citet{NonLinearRedCal} explore another well-known approach, steepest descent, which is much faster per iteration (no matrix inversion is required) but can take many more steps to converge to high precision. The Levenberg-Marquardt algorithm would be a natural compromise between steepest descent and Gauss-Newton, but it and the many other non-linear least-squares minimization techniques are outside the scope of this work.
 

\subsection{Degeneracies in Redundant-Baseline Calibration} \label{sec:1poldegen}

Regardless of the technique for solving for gains and visibilities and regardless of how overdetermined the system in equation~\ref{eq:system} is, the structure of $\chi^2$ (and thus $\mathbf{A}$) guarantees that there will always be a few terms that we cannot solve for with redundant-baseline calibration. These unsolvable quantities are the degeneracies of the system.\footnote{For more concise but less pedagogical discussions of these degeneracies, see \citet{redundant}, \citet{MITEoR}, \citet{ZhengBruteForce}, or \citeauthor{WenyangMWAHex}~(\textit{in~review}).} They manifest as changes that one can make to $g_i$ or $V^\text{sol}_{i-j}$ or both such that $g_i g_j^* V_{i-j}^\text{sol}$ is unchanged, leaving $\chi^2$ also unchanged. Once $\chi^2$ is minimized, any change in these degeneracies keeps $\chi^2$ minimized. In single-polarization calibration, there are exactly four such degeneracies per frequency and per time. They are:
\begin{enumerate}[leftmargin=*,labelindent=10pt, label=\textbf{\arabic*.}]
\item \textbf{The overall amplitude.} If $g_j \rightarrow A g_j$ and $V_{i-j}^\text{sol} \rightarrow V_{i-j}^\text{sol} / A^2$, then $g_i g_j^* V_{i-j}^\text{sol}$ is unchanged.
\item \textbf{The overall phase.} If $g_j \rightarrow g_j e^{i\psi}$, the changes in $g_i$ and $g_j^*$ always exactly cancel out.
\item \textbf{The $x$-phase gradient.} If $g_j \rightarrow g_j e^{i\Phi_x x_j}$ and $V_{i-j}^\text{sol} \rightarrow V_{i-j}^\text{sol} e^{-i \Phi_x \Delta x_{ij}}$, then $g_i g_j^* V_{i-j}^\text{sol}$ is unchanged for all baselines.
\item \textbf{The $y$-phase gradient.} Likewise, if $g_j \rightarrow g_j e^{i\Phi_y y_i}$ and $V_{i-j}^\text{sol} \rightarrow V_{i-j}^\text{sol} e^{-i \Phi_y \Delta y_{ij}}$, then $g_i g_j^* V_{i-j}^\text{sol}$ is similarly unchanged.
\end{enumerate}
The first two degeneracies are due to straightforward cancellations. The third and fourth are a bit more subtle. They rely on the fact one can add a linear phase gradient $\mathbf{\Phi} \equiv (\Phi_X, \Phi_Y)$ in antenna position $\mathbf{r}_j \equiv (x_j, y_j)$ which, due to the complex conjugation of $g_j^*$, produces a phase factor that depends only on baseline $\mathbf{b}_{ij} \equiv (\Delta x_{ij}, \Delta y_{ij}) = (x_i - x_j, y_i - y_j)$. This can be exactly canceled by rephasing the unique visibility solutions, which can depend on baseline vector but not on absolute antenna position. The phase gradient terms are often referred to as the tip-tilt terms since they correspond to moving the phase center and thus the apparent position of sources on the sky \citep{MITEoR,ZhengBruteForce}.

We can think of these four degeneracies as four different vectors in the solution space of the gains and visibilities. Movement along any of these directions does not affect $\chi^2$, so there is no way to know the ``optimal'' amount to move along these directions at each iterative step. This is manifested in the  structure of $\mathbf{A}^\intercal \mathbf{A}$, which has four zero-eigenvalues and a nullspace spanned by the locally linearized versions of these vectors. That said, it is difficult to simply compute the eigenvectors and eigenvalues of $\mathbf{A}^\intercal \mathbf{A}$ and verify that they are the same four vectors by inspection. Eigenvalue decomposition algorithms will produce some unpredictable linear combination of the four degeneracies precisely because they are degeneracies. However, if one adds extra constraints to the system in equation~\ref{eq:system} that fix the degeneracies, $\mathbf{A}^\intercal \mathbf{A}$ becomes full-rank. Since these degeneracies are independent between frequencies, nothing in the $\chi^2$ minimization algorithm requires that they be spectrally smooth.


\subsection{Avoiding Spectral Discontinuities in the Degeneracies} \label{sec:delay_errors}

The immediate question posed by these degeneracies is simply ``what should we do about them?" Ultimately, we need to know how bright the sky is and we need to know where our array is pointed. We will need to fix the degeneracies with a sky-referenced absolute calibration, either through a diffuse sky-model \citep{MITEoR, ZhengBruteForce} or via imaging and traditional self-calibration (\citeauthor{WenyangMWAHex}~\textit{in~review}). 
However, uncertainties about both our sky-model and our beams can lead to enough spectral structure to bias the entire EoR window \citep{ModelingNoise}. Though we need to calibrate four degeneracies per frequency (and time), we will likely want to restrict the degrees of freedom in our absolute calibration so as to avoid adding spectral structure to our overwhelmingly bright foregrounds. If we plan to constrain our absolute calibration in this way, we must ensure that redundant baseline calibration does not add any spectral structure within the degenerate subspace that absolute calibration may not take out.  

Fixing the amplitude degeneracy to achieve this goal is easy; all we have to do is ensure that the degeneracy is constant over frequency and time. There are several ways to do this. We could fix the amplitude of the gain of a particular reference antenna or we could fix the mean amplitude of all antennas. We choose to set the mean amplitude of the gain products (as they appear in equation~\ref{eq:chisq}) to 1. This preserves the mean visibility amplitude, though none of these choices make much difference and or risks adding any spectral structure. 

Phase calibration is trickier. Naively approaching the three phase degeneracies in the same way---for example by setting the mean gain phases and gain phase slopes to zero---is problematized by phase wraps. Antenna gain phases generically evolve with frequency. When one antenna's phase wraps from $\pi$ to $-\pi$, the $\psi$ that we need to add to all phases to make the average 0 might be quite different from one frequency to the next. This creates a phase discontinuity in the gains and in the visibilities. Unless it is fixed by absolute calibration, such a discontinuity in frequency-space will leak foreground power to all modes in Fourier space and will contaminate the EoR window. Working in terms of real and imaginary parts of the gains and visibilities does not solve the phase-wrap problem because the degeneracies are inherently degeneracies in amplitude and phase.

To understand how fixing the degeneracies to a constant value can produce this sort of spectral structure, let us consider separating antenna gains into the form
\begin{equation} \label{eq:delay_formalism}
g_j(\nu) \equiv |g_j(\nu)|e^{i \phi_j(\nu) - 2\pi i \tau_j \nu}
\end{equation}
$\tau_j$ is the delay for the $j$th antenna and $\phi_j$ is its residual phase. In practice, the phase structure of $g_j(\nu)$ is dominated by $\tau_j$ which comes from light-travel time delays along the cable and other delays in the analog and digital signal chain. While $\phi_j$ generally does not phase wrap over the band of interest, typical delays for EoR instruments like MWA, PAPER, and HERA on the order of 10s or 100s of ns over the typical $\sim$100\,MHz bandwidth create many phase-wrappings for each antenna. 

\begin{figure*}
\includegraphics[width=\textwidth]{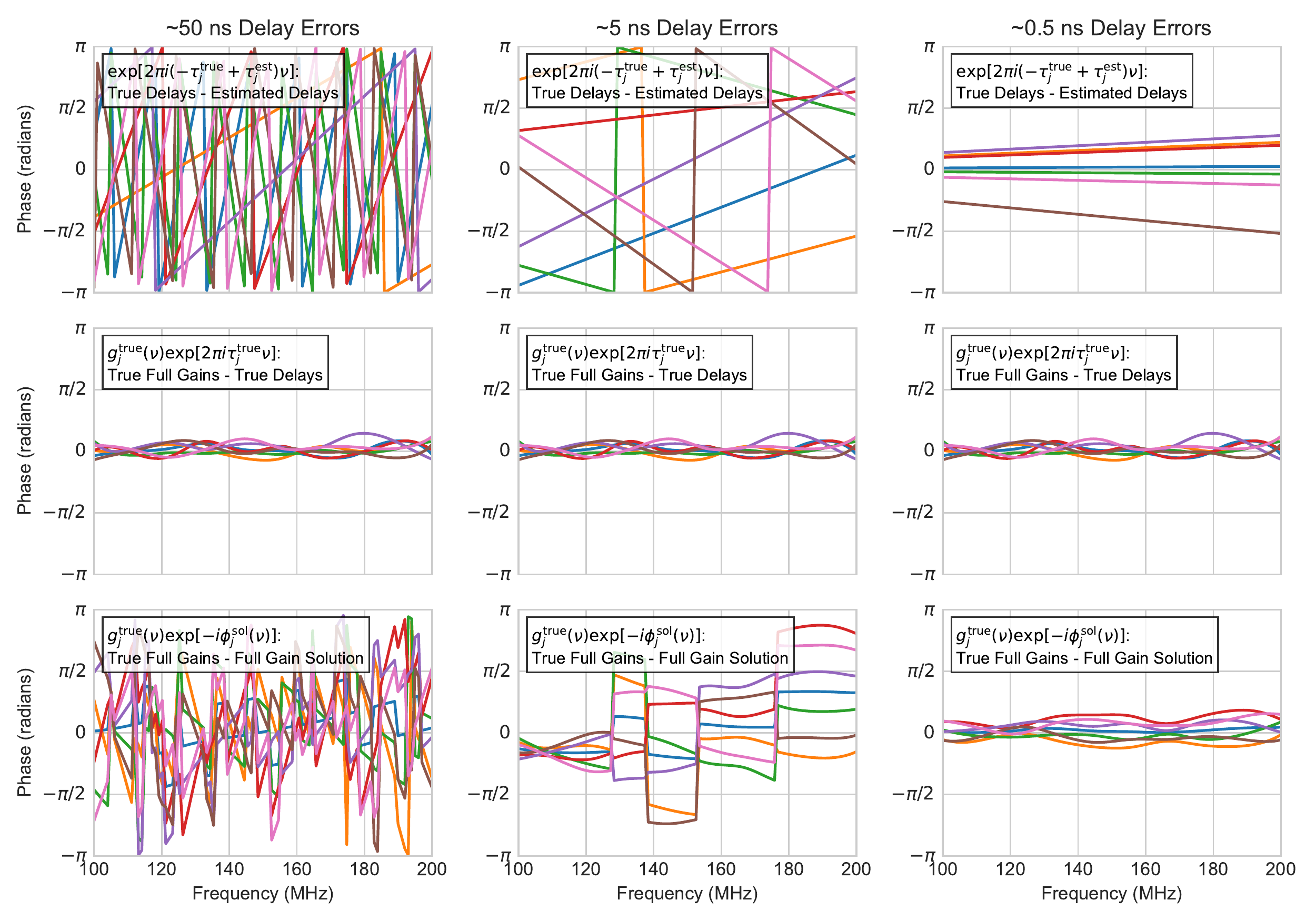}
\caption{Getting accurate delays before performing redundant-baseline calibration is key to avoiding spectral structure introduced by degeneracy-fixing. Here we show the phases of complex antenna gains as a function of frequency for each element in a simulated 7-element HERA observation, comparing the case where the errors in our estimate of the random delays are $\sim$50\,ns, $\sim$5\,ns, and $\sim$0.5\,ns. Each line is a different antenna. Different levels of delay errors produce different amounts of phase wrapping, seen most clearly in the difference between the true delays, $\tau^\text{true}_j$, and the estimated delays, $\tau^\text{est}_j$ (top row). When the true delays are taken out of the true full gains (middle row), the residual phase, $\phi_j$ from equation~\ref{eq:delay_formalism}, is small and simulated to have little intrinsic spectral structure. However, when the true gains are pre-calibated with differing levels of delay errors, the ensuing phase wraps combine with our degeneracy-fixing method to produce discontinuities in the calibration solution if the delay errors are large enough (bottom row). In all three noise-free simulations, the gain and visibility solutions converge to perfect redundancy with the input simulation; redundant-baseline calibration does its job minimizing $\chi^2$. Examining the 5\,ns case, it is clear that individual antenna phase-wraps in the delay error (top row) correspond to and in fact cause dramatic phase jumps in the gain errors (bottom row). The 0.5\,ns case, which does not phase-wrap, still shows some differences between the true gains and the gain solutions, but this is expected. While redundant baseline calibration did not give exact the right answer inside the degenerate subspace, its errors are as spectrally smooth as the simulated gains and therefore much more easily removed with absolute calibration.}
\label{fig:delay_errors}
\end{figure*}

In Figure~\ref{fig:delay_errors} 
we show this effect more clearly for a noise-free simulation of a 7-element version of HERA (14.6\,m separation between 7 dishes arranged in a hexagon). We simulate $\phi_j(\nu)$ with 1\,MHz channels over a fiducial 100\,MHz bandwidth to be small and spectrally smooth (i.e.\ describable by only a few Fourier modes). We also simulate random delays, $\tau^\text{true}_j$, but then pre-calibrate them out imperfectly with estimated delays $\tau^\text{est}_j$:
\begin{equation}
V^\text{pre-cal}_{ij} (\nu) = V^\text{obs}_{ij} (\nu) \exp\left[2\pi i (\tau^\text{est}_{i} - \tau_j^\text{est})\right]. \label{precal}
\end{equation}
To the extent that $\tau^\text{true}_j \approx \tau^\text{est}_j$, equation \ref{precal} reduces to 
\begin{equation}
V^\text{pre-cal}_{ij} (\nu) \approx \left|g_i(\nu)g_j^*(\nu)\right| e^{i\phi_i(\nu) - i\phi_j(\nu)} V^\text{true}.
\end{equation}
In our simulation, however, we intentionally introduce residual delay errors on each antenna at the $\sim$50\,ns, $\sim$5\,ns, or $\sim$0.5\,ns levels. 
These realistically represent the cases where delays are unaccounted for, where delays are corrected for but poorly, and where delays are well-measured and taken out. We then perform redundant baseline calibration, perfectly minimizing $\chi^2$ in all three cases,\footnote{Technically, a noise-free simulation has undefined $\chi^2$, but we simply mean that all the differences in the numerator of equation~\ref{eq:chisq} are zero to within numerical precision.} and show how our decision to fix the three phase degeneracies to zero creates spectral structure in the gain solutions.

Figure \ref{fig:delay_errors} shows us that if we do not do a good job correcting the delays before redundant-baseline calibration and degeneracy-fixing, the difference between the true delays and the delay guesses will phase wrap (Figure~\ref{fig:delay_errors}, top row, left and middle panels), causing phase jumps in all antenna gains (bottom row, left and middle panels). While a few such jumps may be easily spotted, increasing the number of antennas or the delay error level makes this effect so complicated that it becomes difficult to perform sky-referenced absolute calibration.\footnote{\citet{ZhengBruteForce} propose an alternative degeneracy-fixing scheme where three antennas have their phases set to zero. This can ameliorate the problem seen in Figure~\ref{fig:delay_errors}, which presents the worst-case scenario in the left column. However, the alternative approach relies on prior knowledge that the reference antennas are stable and working properly---which may not always be the case. It also is more likely to phase the array far from zenith, making absolute calibration somewhat more difficult.} Assuming that phases are zeroed at the center of the band and phase wraps occur at $\pi$ and $-\pi$, then we need $|\Delta\tau_\text{max}| < ( \nu_\text{max} - \nu_\text{min})^{-1}$. This yields a maximum delay error of 10\,ns for 100\,MHz of bandwidth. The requirement gets more stringent if instead phases have $\phi=0$ at 0\,MHz; in the above example we needed to restrict phase errors to $|\Delta\tau_\text{max}| < 2.5$\,ns to avoid phase wraps, which is conservative but not too terribly challenging.

We cannot just do nothing---we must fix the degeneracies if we want to keep any later degeneracy-fixing to match the sky (i.e. absolute calibration) as simple and smooth as possible. But in doing so, we must make sure we avoid phase-wraps from relatively small delay errors. It follows then we must begin degeneracy-fixing with a smooth, approximate solution. A single delay per antenna is usually sufficient. Since antenna delays are generally stable over time, using the same delays for multiple integrations also gives us consistent degeneracy structure from integration to integration, making absolute calibration easier. Also, since we need a good starting point for linearized redundant-baseline calibration anyway, an accurate per-antenna delay kills two birds with one stone. 

Thankfully, good methods for finding antenna delays already exist, including both a sky-based approach \citep{kern_hera41} and one using only redundant baseline pairs instead of a sky model. This technique, called \texttt{firstcal},\footnote{Also available in the most recent version of \href{https://github.com/HERA-Team/hera_cal/tree/v1.0}{\texttt{hera\_cal}}.} fits the ratio of two redundant visibilities to a single $\tau$ which is a combination of the four antenna $\tau$s involved in the ratio. By building up a system of equations much like that of redundant baseline calibration, $\tau_i$ can be largely solved for. A similar method was used in \citet{PAPER32Limits}, though a more complete and formal description can be found in \citeauthor{WenyangMWAHex}~(\textit{in~review}). This improves upon \citet{MITEoR} which uses a sky and instrument model for so-called ``rough'' calibration. It also improves upon \citet{ZhengBruteForce} which suggests a per-frequency technique that is invulnerable to phase-wrapping but can create much more temporal structure than the more physically-motivated delay approach of \texttt{firstcal}.
There are still three \texttt{firstcal} degeneracies, very similar to the ones in Section~\ref{sec:1poldegen}, and they affect whether source images appear in the right place on the sky. However, since these degeneracies do not source the the kind of spectral structure in the gains that causes the phase-wrapping we are focused on in this work, we forgo a more detailed exploration of them here.


\subsection{Assumptions of Redundant-Baseline Calibration}\label{sec:assumptions}
The need for an accurate delay for each antenna in order to avoid spectral structure highlights the fact that redundant-baseline calibration relies on a number of assumptions to work well. While a single delay per antenna has, in our experience, been close enough to restrict our calibration space to the $-\pi$ to $\pi$ range, there is no \emph{a priori} reason for this to be true. It is possible that we need to initialize our redundant calibration with a low order polynomial in $\nu$, instead just a linear phase slope created by a delay. As long as we can get within a phase wrap, we do not have to worry about degeneracy fixing.

That said, there are many other possible deviations from redundancy. It is beyond the scope of the present work to quantify the impact of these on cosmological measurements. However, our pedagogical review would be incomplete with an explicit enumeration of the possible future issues with redundant baseline calibration:
\begin{itemize}[leftmargin=*,labelindent=10pt]
\item As we already discussed above, it is essential that we can find a spectrally smooth starting solution, ideally a single delay per antenna, that lets redundant-baseline converge to the right answer and prevents degeneracy-fixing from introducing spectral structure into the gain and visibility solutions. 
\item If antennas are not placed perfectly redundantly or if the array is not perfectly coplanar, we break the assumption that nominally redundant baselines are actually measuring the same mode on the sky. Since small model errors can create worrisome spectral structure \citep{ModelingNoise}, this deserves further investigation. However, this effect can potentially be ameliorated by a couple of extensions to standard redundant-baseline calibration framework. If the deviations from the redundant grid are well-measured, one can add additional degrees of freedom that admit such non-redundancy self-consistently by Taylor expanding the visibility solutions as a function of baseline \citep{redundant}. Alternatively, the framework of \citet{CorrCal} incorporates partial redundancy statistically into $\chi^2$, penalizing not-quite redundant baselines less for having differing visibility solutions. Both require further investigation outside the scope of this work.
\item While one generally seeks to make one's elements as identical as possible, some variance from primary beam to primary beam is inevitable. Unfortunately, no method we know of can fully absorb this effect into redundant calibration if the actual beam-to-beam variation is poorly characterized. This is likely to induce spectral structure into calibration as well, as was seen in simulations of sky-based calibration with primary beam errors \citep{ModelingNoise}. This can also manifest as an antenna-to-antenna variation of polarized leakage, which we will discuss in detail in Section~\ref{sec:D-terms}.
\item Correlated noise between two antennas, either due to transmission by one or due to communication between nearby inputs on a printed circuit board, leads to apparent signal that is not redundant between baselines \citep{kohn_hera6}. This crosstalk can be mitigated by good antenna design \citep{HERAOverview}, in the electronics with Walsh modulation \citep{MITEoR}, or by filtering signals that apparently do not change as the sky rotates \citep{FringeRateFilter}, but the exact magnitude of the effect remains an open question.
\item Pairs of nominally redundant baselines that are separated from each other by several km see a different isoplanetic patch of the ionosphere \citep{HarishLeonIon1}, leading to two slightly differently warped skies and thus different visibilities. At these distances we might also begin to worry about array non-co-planarity and pointing errors due to the curvature of the Earth. We include this effect for completeness; it is not expected to affect the redundant-baseline calibration of any operational or planned highly-redundant interferometer.
\end{itemize}


\section{Incorporating Polarization in Redundant-Baseline Calibration}\label{sec:polarization}

Having summarized the challenges presented by the challenges of redundant-baseline calibration, we now are ready relax the simplification that we only need to calibrate a single antenna polarization. Most 21\,cm interferometers (including LOFAR, GMRT, MWA, PAPER, CHIME, HIRAX, HERA and SKA-low) are dual-polarization instruments. They simultaneously measure electromagnetic signals from two orthogonal antenna polarizations, $n$ and $e$.\footnote{These are often referred to $x$- and $y$-polarizations (to highlight their orthogonality), but we use $e$ and $n$ (for East-West and North-South orientations) to avoid confusion with antenna position. Of course, there is no requirement that the polarizations must line up with the cardinal directions, but we are assuming that they are orthogonal to one another.} These are correlated to form four visibility polarizations, $V^{ee}_{ij}$, $V^{en}_{ij}$, $V^{ne}_{ij}$, and $V^{nn}_{ij}$. The cross-polarized visibilities, $V^{en}_{ij}$ and $V^{ne}_{ij}$, will generally have much lower signal-to-noise (SNR) ratios than the parallel-polarized visibilities, $V^{ee}_{ij}$ and $V^{nn}_{ij}$, though they have the same noise levels. This is because the cross-polarized visibilities are only sensitive to intrinsically linearly-polarized emission ($<$1\% of the foregrounds) and leakage from unpolarized emission ($\sim$10\% of the foregrounds near the horizon) which usually dominates in 21\,cm cosmology \citep{EmilPol,SaulPAPERPol,NunhokeePolarized}. However, all four visibility polarizations are necessary to make images of each of the four Stokes parameters $I$, $Q$, $U$, and $V$ \citep{ThompsonMoranSwenson}. The generalization to polarized redundant-baseline calibration has not been addressed in the literature and presents new and unique challenges.

Generalizing equation~\ref{eq:system}, we now model our observations as
\begin{equation} \label{eq:system_4pol}
V_{ij}^{ab,\text{obs}} = g^a_i g^{b*}_j V_{i-j}^{ab,\text{sol}}
\end{equation}
where $a$ and $b$ stand in for either $e$ or $n$ for antennas $i$ and $j$ respectively and we have again omitted the explicit frequency dependence.\footnote{Though for now we assume that a single complex gain per antenna is sufficient, we will relax this assumption in Section~\ref{sec:D-terms}.} If we only want to calibrate the highest SNR visibilities---$V^{ee}_{ij}$ and $V^{nn}_{ij}$---then this problem decouples into the independent calibration of the East-West visibilities with the East-West gains and the North-South visibilities with the North-South gains. Degeneracy fixing is the same as in Section~\ref{sec:1poldegen}, effectively leaving the problem of relative calibration between $e$ and $n$ for later absolute calibration. 

Though we have four times as many visibilities, we only have double the number of complex gains; this gives us a new way to connect together visibilities self-consistently via the gains. Ideally, we would solve for everything simultaneously, minimizing a $\chi^2$ generalized from equation~\ref{eq:chisq} to 
\begin{equation} \label{eq:chisq_4pol}
\chi^2= \sum_{a,b \in e,n} \left[ \sum_{\text{all pairs }i,j} \frac{\left|V_{ij}^{ab,\text{obs}} - g^a_i g_j^{b*} V^{ab,\text{sol}}_{i-j}\right|^2}{\left(\sigma^{ab}_{ij}\right)^2} \right],
\end{equation}
This couples all four visibility polarizations via the antenna gains into one large system of equations. It produces a very similar $\mathbf{A}$ matrix to that described in equations~\ref{eq:real} and \ref{eq:imag}, but with four times as many equations, nearly four times as many parameters to solve for.\footnote{An explicit derivation of $\mathbf{A}$ for full-polarization $\chi^2$ minimization is not particularly illuminating and so we omit it here.}


\subsection{Degeneracies in Polarized Redundant-Baseline Calibration} \label{sec:4poldegen}

Just as in the single polarization case, the form of $\chi^2$ in equation~\ref{eq:chisq_4pol} features several different ways to modify the gains and visibilities so as to leave $\chi^2$ unchanged. Once again, these degeneracies are key to maintaining spectral smoothness and depend on precisely what analytical approach we take. 

The polarized extension of redundant-baseline calibration where we only consider the high SNR parallel-polarized visibilities, $V^{ee}_{ij}$ and $V^{nn}_{ij}$, ignoring $V^{en}_{ij}$ and $V^{ne}_{ij}$, is trivial. The $e$ and $n$ polarizations decouple and there are precisely eight degeneracies---the same four we saw in Section~\ref{sec:1poldegen} for each of the two antenna polarizations. This approach, which we refer to as \emph{2-pol} calibration (in contrast \emph{1-pol} above or \emph{4-pol} below), is simpler and computationally cheaper, but comes at the cost of throwing away some information. The cross-polarized visibilities are left to be calibrated later using the gains from the 2-pol solution. 

The best approach in an ideal world would be the minimize $\chi^2$ for both gain polarizations and all four visibility polarizations simultaneously. Decomposing the associated $\mathbf{A}$ matrix reveals that the nullspace is spanned by six eigenvectors, not eight. Bringing in $V^{en}_{ij}$ and $V^{ne}_{ij}$ apparently allows us to solve for two of the eight degeneracies. In retrospect, this is not surprising as there are six independent ways that gains and visibilities can be modified without changing $\chi^2$ in equation~\ref{eq:chisq_4pol}. Those 4-pol degeneracies are:
\begin{enumerate}[leftmargin=*,labelindent=10pt, label=\textbf{\arabic*.}]
\item \textbf{The overall $e$-polarization amplitude.} All $g^e_j \longrightarrow A_e g^e_j$.
\item \textbf{The overall $n$-polarization amplitude.} All $g^n_j \longrightarrow A_n g^n_j$. The combined effect of these two can be perfectly canceled by transforming the visibility solutions as:
\begin{equation}
V_{i-j}^{ab,\text{sol}} \rightarrow V_{i-j}^{ab,\text{sol}} / (A_a A_b)
\end{equation}
where both $a$ and $b$ stand in for either $e$ or $n$.
\item \textbf{The $e$-polarization overall phase.} All $g_j^e \longrightarrow g_j^e e^{i\psi_e}$.
\item \textbf{The $n$-polarization overall phase.} All $g_j^n \longrightarrow g_j^n e^{i\psi_n}$. Once again, these effects can be canceled by transforming the visibility solutions as:
\begin{equation}
V_{i-j}^{ab,\text{sol}} \rightarrow V_{i-j}^{ab,\text{sol}} e^{i(\psi_a - \psi_b)}.
\end{equation}
Here $V_{i-j}^{nn}$ and $V_{i-j}^{ee}$ are unmodified, as in the 1-pol case.
\item \textbf{The $x$-phase gradient.} Just as in the 1-pol case, if $g^a_j \rightarrow g^a_j e^{i\Phi_x x_j}$ and $V_{i-j}^{ab,\text{sol}} \rightarrow V_{i-j}^{ab,\text{sol}} e^{-i \Phi_x \Delta x_{ij}}$, then $\chi^2$ is unchanged.
\item \textbf{The $y$-phase gradient.} Likewise, if $g^a_j \rightarrow g^a_j e^{i\Phi_y y_j}$ and $V_{i-j}^{ab,\text{sol}} \rightarrow V_{i-j}^{ab,\text{sol}} e^{-i \Phi_y \Delta y_{ij}}$, then $\chi^2$ is unchanged.
\end{enumerate}

It is notable that the last two degeneracies are polarization-independent. By introducing $V_{ij}^{en}$ and $V_{ij}^{ne}$, we have broken the two independent phase gradient degeneracies that we get in the 2-pol case. To understand this, consider the alternative case where $g_i^e \rightarrow g^e_i e^{i\Phi^e_x x_i}$ and  $g_j^n \rightarrow g^n_j e^{i\Phi^e_x x_j}$. We would need to transform $V_{i-j}^{en} \longrightarrow V_{i-j}^{en} e^{-i (\Phi^e_x x_i - \Phi^n_x x_j)}$. If $\Phi^e_x \neq \Phi^n_x$, then the phase factor cannot be factored in the exponent and the new visibility becomes an explicit function of antenna position, not just the baseline vector. But, assuming antennas are identical and perfectly positioned, the unique visibility $V_{i-j}^{en}$ cannot depend on position. This contradiction is only resolved only if $\Phi^e_x = \Phi^n_x = \Phi_x$ and $\Phi^e_y = \Phi^n_y = \Phi_y$.


\subsection{Difficulties of 4-Polarization Redundant-Baseline Calibration}

Breaking two additional degeneracies---and therefore having to rely on absolute calibration to solve for two fewer numbers per frequency and time---sounds really useful. Unfortunately, it is not so simple. Cross-polarized visibilities, as we mentioned above, have much lower SNR than parallel-polarized visibilities. Since introducing $V_{ij}^{en}$ and $V_{ij}^{ne}$ broke two of the four phase gradient degeneracies, it follows that our constraint within that subspace comes only from cross-polarized observations. While we have information to break the degeneracy, it is very noisy compared to the rest of the data that informs our calibration.

\begin{figure}
\includegraphics[width=.49\textwidth]{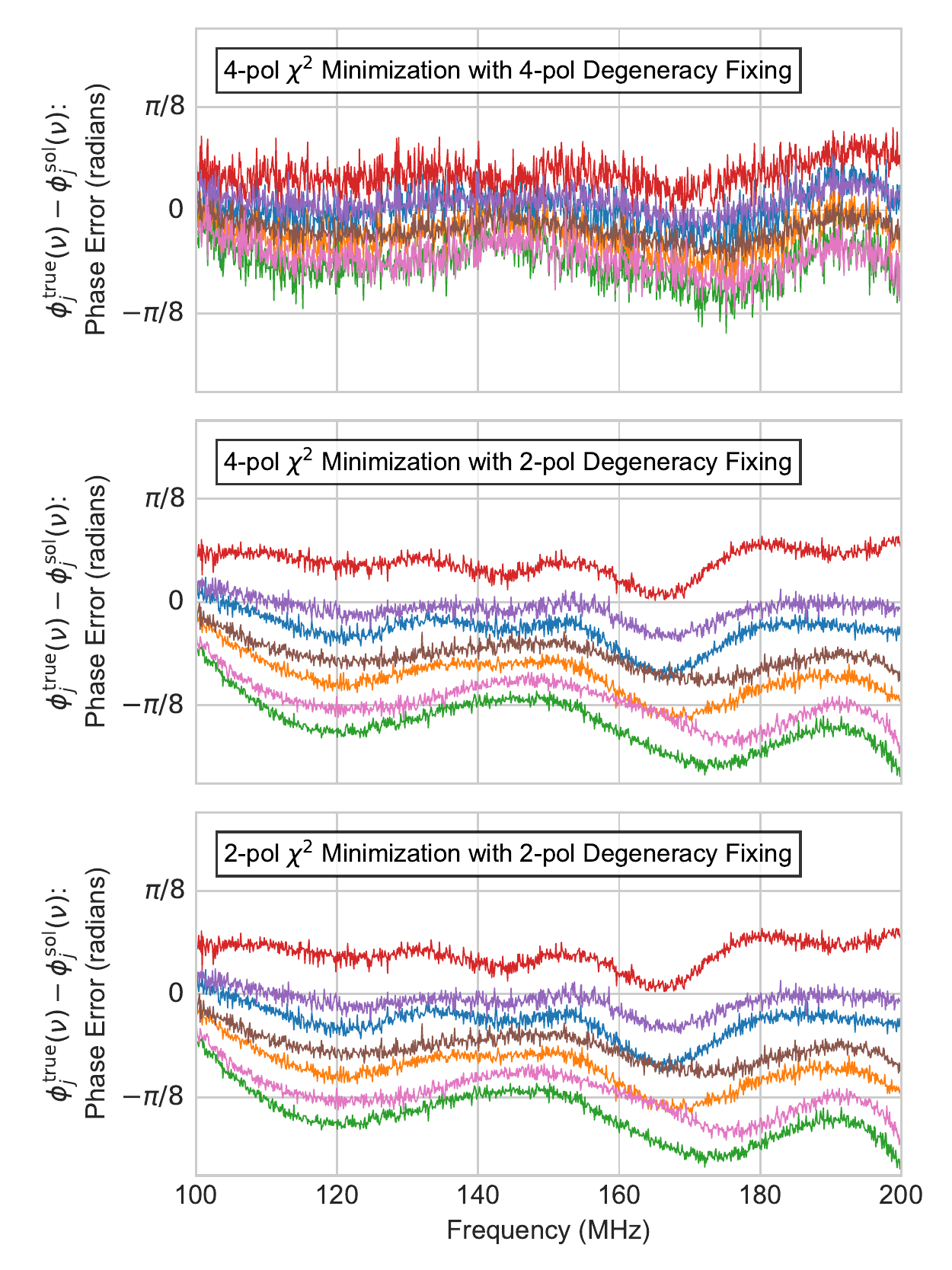}
\caption{While 4-pol calibration has the potential to break two of the eight redundant-baseline calibration degeneracies, these are only broken by the addition of cross-polarized visibilities that tie $e$ and $n$ polarizations together. These visibilities generally have much less sky-signal, which means that they have low SNR. Fixing only the six true degeneracies described in Section~\ref{sec:4poldegen} to 1 (for amplitudes) or 0 (for phases) gives smaller errors on the gains at the cost of much nosier gain solutions. Here we plot the error in the phase of a subset of the antenna gains of a 19-element array with three different calibration techniques. First, we show the result of 4-pol redundant baseline calibration after fixing the six degeneracies (top panel). Next we show the result of the same $\chi^2$ minimization procedure but after fixing the eight degeneracies of 2-pol calibration (middle panel). This essentially throws out the information in those degenerate modes, saving the problem for later absolute calibration. It produces very similar results to having simply excluded the cross-polarized visibilities from both $\chi^2$ minimization and degeneracy-fixing (bottom panel).}
\label{fig:4_pol_RD}
\end{figure}

The effect of this is shown most clearly in Figure~\ref{fig:4_pol_RD}. 
We simulate spectrally smooth gains and visibilities for a 19-element HERA-like array, much like we did in Figure~\ref{fig:delay_errors} (without delay errors), but now for both $e$ and $n$ antenna polarizations and all four visibility polarizations.  Then we add white noise at equal levels to all ``observed'' visibilities, but only after decreasing the signal level in the cross-polarized visibilities by a factor of 10. This gives our parallel-polarized visibilities an SNR of $\sim$10 and our cross-polarized visibilities an SNR of $\sim$1. 

We then assess three different calibration techniques and plot the gain phase errors they produce. The first is standard 4-pol redundant-baseline calibration (top panel). The second is standard 2-pol calibration (bottom panel). The third is a hybrid approach where we minimize $\chi^2$ as in the 4-pol approach with all four visibility polarizations, but then throw away $V^{en}_{i-j}$ and $V^{ne}_{i-j}$ and fix the eight degeneracies of the 2-pol case (middle panel).

The 4-pol approach has smaller phase errors than the other two, but it is much noisier. The additional constrained degrees of freedom in the 4-pol approach, the two phase gradient degeneracies, are only informed by the low SNR cross-polarized visibilities. No inverse noise variance weighting scheme can help us here. It is not that the high-SNR parallel-polarized visibilities are given too little weight relative to the low-SNR cross-polarized visibilities. The parallel-polarized visibilities get zero weight in the determination of these two modes, regardless of how much we up-weight them, because they contain no information about the modes. The rest of the $\chi^2$ landscape is dominated by the parallel-polarized visibilities, even when they get equal noise weighting, because they have much larger signals. The extra noise effect goes away if cross-polarized and parallel polarized visibilities have the same SNR. Likewise, the improvement in the phase errors goes away as the signal level in the cross-polarized visibilities goes to zero.

The smaller errors are unlikely to be worth the additional cost. This extra noise gets spread through the gains to the calibrated parallel-polarization visibilities, giving them a much higher effective noise level that will integrate down more slowly with repeated observations. It seems better to effectively impose a strong prior that we trust sky-based absolute calibration for these modes---which ultimately correspond to the relative pointing of the array between $n$ and $e$---in order to not introduce additional spectral structure into the gains and high SNR visibilities. 

It may be possible to extract a constrained or smoothed solution for the two additional degeneracies that has many fewer degrees of freedom and thus is less noisy. Doing so will require additional assumptions about true relative phase gradients and so we leave that investigation to future work. 

Likewise, one possible instrumental solution would be to rotate some of the antenna feeds by $\pm 45\degree$, producing a dipole pattern that looks, for example, like $+$~$\times$~$+$~$\times$ from above. We would then have four dipole orientations, $e$, $n$, $+$, and $-$, and up to sixteen visibility polarizations per baseline. While cross-polarized visibilities have very low SNR, visibilities like $V_{ij}^{e+}$ should actually exhibit only a moderate hit to SNR. This could connect together all visibility in one redundant-calibration system through relatively high SNR measurements. On the other hand, it increases the complexity of the calibration and mapping problems and reduces the number of simultaneously redundant baselines for any particular separation. Not all array configurations that are redundant with two dipole orientations polarizations are still redundant when some of the antennas are rotated. The possibility merits further investigation, though because no current or planned experiment features a mix of $e$/$n$ and $+$/$-$ orientated feeds, we again leave it for future work.

This all raises another question: is it worthwhile to bother including the cross-polarized visibilities in $\chi^2$ minimization at all? It is certainly more computationally difficult to perform fully-polarized redundant baseline calibration. By doubling the number of gains and quadrupling the number of visibilities solved for simultaneously, we increase the cost of matrix inversion in equation~\ref{eq:noiseInvEstimator} by up to a factor of 64, though if we only have to perform the inversion half as often as in the 2-pol case. However, if our chief aim is to make a measurement of polarized foregrounds, then excluding cross-polarized visibilities from $\chi^2$ will generally lead to calibrated cross-polarized visibilities that are less redundant with each other---effectively a noisier measurement. Even if all we care about is parallel-polarized visibilities for a PAPER-style power spectrum analysis \citep{PAPER32Limits}, we should expect that including cross-polarized visibilities leads to lower noise. More information is always better, even if it is noisy. 

That decrease in noise is real, though difficult to see in Figure~\ref{fig:4_pol_RD}. It turns out to be quite small in our simulation. Comparing the gains only in the non-degenerate subspace of 2-pol redundant calibration, we find that $\chi^2$ minimization with all four visibility polarizations decreases the errors in the gains by $0.62\%$. This improvement is small for a number of reasons: the number of data points doubled, but the new ones all had low SNR. The number of new variables also increased for the cross-polarized visibility solutions, though it did not quite double. The improvement in the parallel-polarized visibility solutions, only $0.0044\%$, is roughly the fractional improvement squared because cross- and parallel-polarized visibilities are connected only indirectly in the linear system in equation~\ref{eq:system_4pol}. Apparently, as long as the assumption of redundancy holds, we need to choose between a small decrease in noise and a sizable increase in computational cost. 


\subsection{The Effect of Non-Redundant Polarization Leakage} \label{sec:D-terms}

Another concern about 4-pol $\chi^2$ minimization is the possibility that cross-polarized visibilities may be less redundantly-calibratable than parallel-polarized visibilities. The sources of non-redundancy like position errors and beam-to-beam variation, which we discussed in Section~\ref{sec:assumptions}, affect both types of visibilities in similar ways. This is not the case with antenna polarization leakage, often referred by its symbol as $D$\emph{-terms}.\footnote{Two different effects are called ``polarization leakage,'' and it is important to be clear what we mean. Our earlier discussions of visibility polarization leakage (Sections~\ref{sec:intro} and \ref{sec:polarization}) were focused on leakage from Stokes $I$ into cross-polarized visibilities and from Stokes $Q$ and $U$ into the parallel-polarized visibilities. These effects are largely due to beam geometry and are biggest near the horizon for a zenith-pointed instrument. They peak at $\sim$10\% for PAPER \citep{NunhokeePolarized}, through they can vary substantially between feed and element designs. In this paper, we mean only the direction-independent polarization leakage terms described by Equation~\ref{eq:D-terms}.} $D$-terms represent the complex response of the $e$ or $n$ antenna feed to the $n$- or $e$-polarized component, respectively, of the local electric field \citep{RadioPolarimetry1,RadioPolarimetry2}. They are also another way for Faraday rotation to produce spectral structure in our Stokes~$I$ estimate due to varying Stokes~$U$ and Stokes~$V$ leakage as a function of frequency.

Following \citet{ThompsonMoranSwenson}, we can write the observed visibilities in terms of the ``true'' visibilities as
\begin{align}
\frac{V^{ee, \text{obs}}_{ij}}{g_i^e g_j^{e*}} &= V^{ee}_{ij} + D_{i}^{n\rightarrow e} V^{ne}_{ij} + D_{j}^{n\rightarrow e *} V^{en}_{ij} + D_{i}^{n\rightarrow e}D_{j}^{n\rightarrow e *} V^{nn}_{ij} \nonumber \\
\frac{V^{en, \text{obs}}_{ij}}{g_i^e g_j^{n*}} &= V^{en}_{ij} + D_{i}^{n\rightarrow e} V^{nn}_{ij} + D_{j}^{e\rightarrow n *} V^{ee}_{ij} + D_{i}^{n\rightarrow e} D_{j}^{e\rightarrow n *} V^{ne}_{ij} \nonumber \\
\frac{V^{ne, \text{obs}}_{ij}}{g_i^n g_j^{e*}} &= V^{ne}_{ij} + D_{i}^{e\rightarrow n} V^{ee}_{ij} + D_{j}^{n\rightarrow e *} V^{nn}_{ij} + D_{i}^{e\rightarrow n} D_{j}^{n\rightarrow e *} V^{en}_{ij} \nonumber \\
\frac{V^{nn, \text{obs}}_{ij}}{g_i^n g_j^{n*}} &= V^{nn}_{ij} + D_{i}^{e\rightarrow n} V^{en}_{ij} + D_{j}^{e\rightarrow n *} V^{ne}_{ij} + D_{i}^{e\rightarrow n} D_{j}^{e\rightarrow n *} V^{ee}_{ij}, \label{eq:D-terms}
\end{align}
where, for example, $D_{j}^{n\rightarrow e}$ represents leakage from intrinsic $n$-polarized electric field to $e$-oriented antenna feed on the $i$th antenna. $D$-terms are generally small ($\sim$1\%), but their effect on cross-polarized visibilities can be important. They provide a mechanism by which parallel-polarized visibilities with $\sim$10 times larger magnitudes can significantly affect the cross-polarized antennas.

While the presence of $D$-terms makes polarized mapmaking more challenging, it is not inherently a problem for redundant calibration. If the $D$-terms were all identical from antenna to antenna, we could simply redefine our ``true'' visibilities to be the right-hand side of equation~\ref{eq:D-terms} and solve for model visibilities that are actually some unknown linear combination of all four visibility polarizations. In practice, that is not the case. 

\begin{figure}
\includegraphics[width=.49\textwidth]{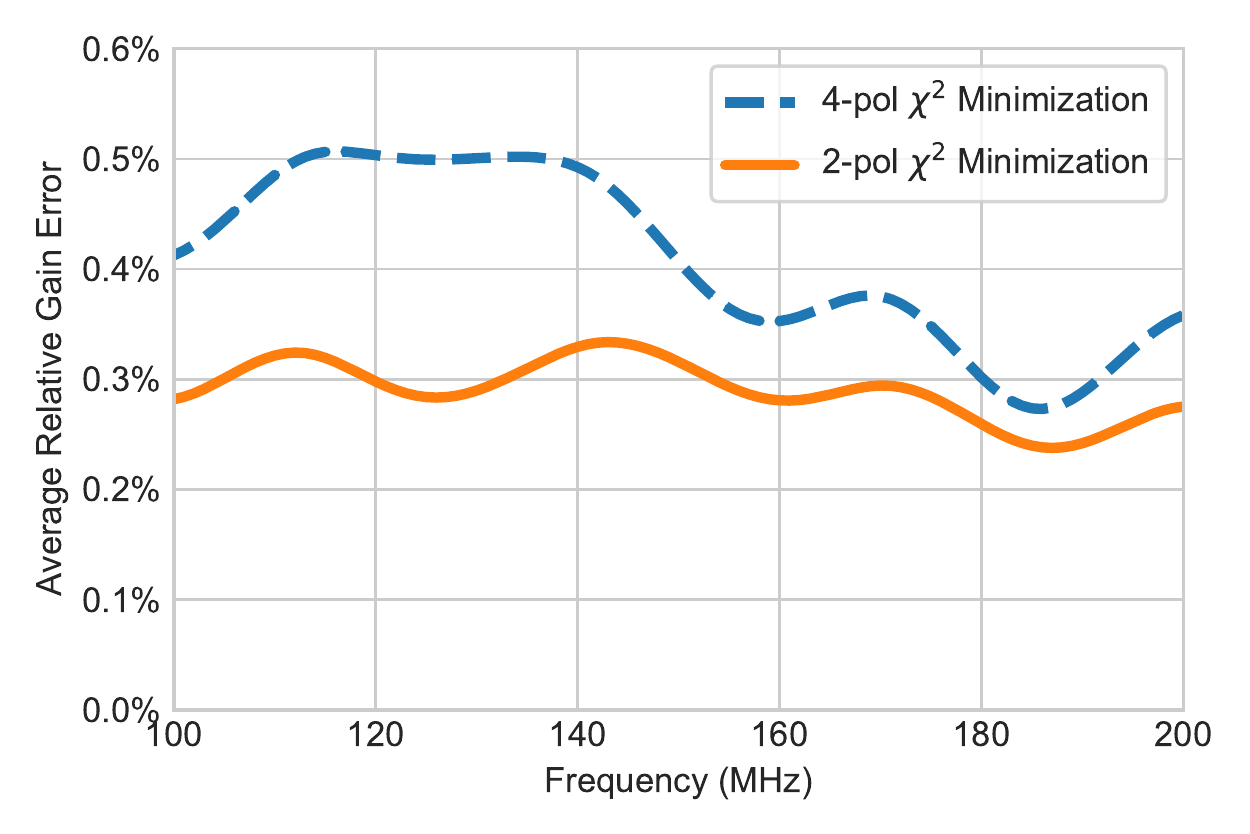}
\caption{Antenna polarization leakage due to the $\sim$1\% sensitivity of an antenna to a perpendicularly polarized electric field, known as $D$-terms, is a potentially important source of non-redundancy. Here we see the resulting average relative gain errors (as defined in equation~\ref{eq:error_metric}) from a simulation of a HERA-19-like array with smooth spectrum gains, visibilities, and $D$-terms, much like in Figures~\ref{fig:delay_errors} and \ref{fig:4_pol_RD}. We calibrate our simulation using only parallel-polarized visibilities (2-pol) and with both parallel- and cross-polarized visibilities (4-pol). Both calibrations produce gain errors because antenna-to-antenna variation in $D$-terms breaks the assumption of redundancy that underlies equation~\ref{eq:system_4pol}. However, because cross-polarized visibilities generally have lower magnitudes than parallel-polarized, $D$-term leakage biases cross-polarized visibilities more than parallel-polarized visibilities. This makes them less redundant, leading to larger errors in the 4-pol $\chi^2$ minimization approach.}
\label{fig:D_terms}
\end{figure}

To explore the effect of $D$-terms that vary from antenna to antenna, we again simulate a noise-free HERA-like 19-element array with relatively smooth foregrounds, gains, and $D$-terms. Gains are set to have an average value of 1.0 with random, antenna-varying but spectrally smooth real and imaginary components at the 15\% level. Likewise, $D$-terms have an average of zero but have a random, smooth component that varies from antenna to antenna the at the 1\% level. Our cross-polarized visibilities have lower signal amplitudes by $\sim$10 than our parallel polarized visibilities. To compare the results from  2-pol and 4-pol $\chi^2$ minimization, we first take the gains, $g_i^a$ coming out of both algorithms and fix all eight 2-pol degeneracies. Next we replace those degeneracies with the true simulated gains to produce $\bar{g}_j^{a}$ and compute gain errors for each antenna as a function of frequency. Due to our degeneracy-fixing and replacement, the errors are only in the non-degenerate subspace of 2-pol redundant-baseline calibration and are due solely to $D$-terms, not noise. Finally, in Figure~\ref{fig:D_terms}, we plot the average of those errors defined as
\begin{equation} \label{eq:error_metric}
\varepsilon = \frac{\sum_{a \in e,n} \sum_{j} \left|\bar{g}_j^{a} - g_j^{a,\text{true}}\right| }{\sum_{a \in e,n} \sum_{j} g_j^{a,\text{true}}}.
\end{equation}

Both methods have identical degeneracy-fixing, substituting the degenerate subspace of the true gain solutions into the eight degenerate degrees of freedom, in order to make the errors directly comparable. Without $D$-terms, both techniques would produce the exact correct answer. As expected, introducing $D$-terms on the order of $\sim$1\% produces comparable errors in the gain estimates, generally less than 0.5\%, independent of array size. The simple redundant calibration model from equation~\ref{eq:system_4pol} that we are optimizing cannot possibly perfectly reproduce the observed visibilities from equation~\ref{eq:D-terms} when the $D$-terms vary. The precise magnitude of the error should not be taken too literally since the average magnitude of $D$-terms depends on antenna design choices, like using a dish or a dipole, and on the precision of element construction.

Interestingly, the effect is about 40\% worse when using 4-pol $\chi^2$ minimization compared to 2-pol $\chi^2$ minimization, even with the same degeneracy-fixing method. This makes sense: $D$-terms leak more signal from parallel-polarized visibilities into cross-polarized visibilities than vice versa precisely because there is more signal to leak. This means that $D$-terms make cross-polarized visibilities less redundant than their parallel-polarized counterparts and thus the assumptions underlying 4-pol redundant-baseline calibration weaker. Thankfully the errors do not introduce additional spectral structure beyond what was already in the simulation, though 4-pol calibration clearly magnifies the error associated with ignoring $D$-terms compared to the the 2-pol approach. 

It may be possible to simultaneously calibrate $D$-terms, gains, and visibilities through a generalized redundant-baseline calibration formalism based on equation~\ref{eq:D-terms}. It is not clear if $D$-terms are the leading-order deviation from non-redundancy and therefore the best way to extend the degrees of freedom in the calibration model. It is also likely that such a generalization will introduce additional degeneracies which will need to fix in a frequency-smooth way. We leave those questions for future work. 


\section{Discussion of Analysis Choices for Redundant-Baseline Calibration}

In this paper we explored some of the subtleties of calibrating antenna gains and phases by taking advantage of repeated, supposedly identical measurements. Using redundant baselines, one can solve for most---but not all---of the calibration degrees of freedom. These remaining unsolved quantities, the \emph{degeneracies} of redundant-baseline calibration, must be resolved using information about the sky. Redundant baseline calibration makes the problem of subsequent absolute calibration much easier. Instead of calibrating one complex number per polarization per antenna per frequency, one only needs to calibrate six to eight numbers per frequency for the whole array, depending on one's analysis approach. Good calibration is key to preserving the separation in Fourier space of the 21\,cm signal and the spectrally smooth foregrounds that are $\sim$$10^5$ times brighter. Small errors can make a big difference.

Our strategy for dealing with the redundant calibration degeneracies focuses on simplifying the later absolute calibration that will ultimately resolve them. In principle, any solution that redundant calibration produces in the degenerate subspace can be removed and recalibrated. In practice, we will likely want to restrict the degrees of freedom that go into absolute calibration to avoid adding spectral structure on scales we have an \emph{a priori} reason to believe that the instrument's spectral response is smooth. Enabling these sorts of restrictions on absolute calibration means that we must be very careful to avoid adding extra spectral structure in the degeneracies. We saw in Section~\ref{sec:delay_errors} how large delay errors could cause exactly such a problem and showed the accuracy necessary to avoid introducing spectral structure.

We also explored the generalization of redundant-baseline calibration to observations with both orthogonal antenna polarizations (and thus four visibility polarizations). We derived the new degeneracies in Section~\ref{sec:4poldegen}, showing how the expansion to two antenna polarizations doubles the number of degeneracies to eight per frequency, and how the inclusion of cross-polarized visibilities breaks two of those eight. However, because cross-polarized visibilities generally have much lower SNR than the parallel-polarized visibilities, that degeneracy breaking is noisy and risks adding spectral structure that will integrate down more slowly that the rest of the thermal noise (see Figure~\ref{fig:4_pol_RD}).

This problem can be solved in two ways. Either we simply ignore the cross-polarized visibilities and calibrate the instrumental polarizations completely separately, the 2-pol approach, or we calibrate all four visibilities together (4-pol $\chi^2$ minimization) but ignore the two weakly broken degeneracies (2-pol degeneracy fixing). The former approach throws away information, raising noise slightly, and it can lead to less redundant calibrated cross-polarized visibilities. The later approach also solves the pressing problem of high noise across the entire band and produced slightly less noisy calibration solutions, but it introduces extra vulnerability to the antenna-to-antenna variation of polarization leakage $D$-terms. These non-redundancies break the assumptions of redundant-baseline calibration asymmetrically, biasing the cross-polarized visibilities more strongly than the parallel-polarized visibilities.

For cosmological measurements, the 2-pol approach seems to be most conservative approach for now. For polarized foreground studies, the case may be different. The calibration simulations in this work are physically motivated, but are still somewhat simplistic. Further simulation with realistic beams is needed to quantify the difference between the 2-pol and 4-pol approaches. It may be that the right answer is not the same for every instrument. The 2-pol approach also happens to be the cheapest, computationally. 

That said, much work remains to make redundant-calibration robust to deviations from redundancy due to position errors, beam variations, $D$-terms, etc. Perhaps the question of polarization will be worth revisiting in the context of techniques for incorporating or mitigating non-redundancy (e.g.\ \citealt{redundant} or \citealt{CorrCal}). As our instrumental knowledge and the sophistication of our calibration methods improves, we may be able to take advantage of the extra information and degeneracy breaking power of the cross-polarized visibilities that we lose by conservatively excluding them from redundant-baseline calibration.


\section*{Acknowledgements}
This work is supported by the National Science Foundation under grants \#1440343 and \#1636646, the Gordon and Betty Moore Foundation, and with institutional support from the HERA collaboration partners. JSD gratefully acknowledges the support of the NSF AAPF award \#1701536 and the Berkeley Center for Cosmological Physics. SAK is supported by the University of Pennsylvania School of Arts and Sciences' Dissertation Completion Fellowship. ARP acknolwedges support of NSF CAREER award \#1352519. ARP and AL acknowledge support from the University of California Office of the President Multicampus Research Programs and Initiatives through award MR-15-328388 as part of the University of California Cosmic Dawn Initiative. GB acknowledges support from the Royal Society and the Newton Fund under grant NA150184. This work is based on research supported in part by the National Research Foundation of South Africa (grant No. 103424). WL and JCP acknowledge the support from NSF grant \#1613040. AL acknowledges support for this work by NASA through Hubble Fellowship grant \#HST-HF2-51363.001-A awarded by the Space Telescope Science Institute, which is operated by the Association of Universities for Research in Astronomy, Inc., for NASA, under contract NAS5-26555.

\bibliographystyle{mnras}
\bibliography{refs}

\begin{thebibliography}{}
\makeatletter
\relax
\def\mn@urlcharsother{\let\do\@makeother \do\$\do\&\do\#\do\^\do\_\do\%\do\~}
\def\mn@doi{\begingroup\mn@urlcharsother \@ifnextchar [ {\mn@doi@}
  {\mn@doi@[]}}
\def\mn@doi@[#1]#2{\def\@tempa{#1}\ifx\@tempa\@empty \href
  {http://dx.doi.org/#2} {doi:#2}\else \href {http://dx.doi.org/#2} {#1}\fi
  \endgroup}
\def\mn@eprint#1#2{\mn@eprint@#1:#2::\@nil}
\def\mn@eprint@arXiv#1{\href {http://arxiv.org/abs/#1} {{\tt arXiv:#1}}}
\def\mn@eprint@dblp#1{\href {http://dblp.uni-trier.de/rec/bibtex/#1.xml}
  {dblp:#1}}
\def\mn@eprint@#1:#2:#3:#4\@nil{\def\@tempa {#1}\def\@tempb {#2}\def\@tempc
  {#3}\ifx \@tempc \@empty \let \@tempc \@tempb \let \@tempb \@tempa \fi \ifx
  \@tempb \@empty \def\@tempb {arXiv}\fi \@ifundefined
  {mn@eprint@\@tempb}{\@tempb:\@tempc}{\expandafter \expandafter \csname
  mn@eprint@\@tempb\endcsname \expandafter{\@tempc}}}

\bibitem[\protect\citeauthoryear{{Ali} et~al.,}{{Ali}
  et~al.}{2015}]{PAPER64Limits}
{Ali} Z.~S.,  et~al., 2015, \mn@doi [\apj] {10.1088/0004-637X/809/1/61}, \href
  {http://adsabs.harvard.edu/abs/2015ApJ...809...61A} {809, 61}

\bibitem[\protect\citeauthoryear{{Asad} et~al.,}{{Asad}
  et~al.}{2015}]{AsadPolarization1}
{Asad} K.~M.~B.,  et~al., 2015, \mn@doi [\mnras] {10.1093/mnras/stv1107}, \href
  {http://adsabs.harvard.edu/abs/2015MNRAS.451.3709A} {451, 3709}

\bibitem[\protect\citeauthoryear{{Asad}, {Koopmans}, {Jeli{\'c}}, {de Bruyn},
  {Pandey}  \& {Gehlot}}{{Asad} et~al.}{2018}]{AsadPolarization3}
{Asad} K.~M.~B.,  {Koopmans} L.~V.~E.,  {Jeli{\'c}} V.,  {de Bruyn} A.~G.,
  {Pandey} V.~N.,   {Gehlot} B.~K.,  2018, \mn@doi [\mnras]
  {10.1093/mnras/sty258}, \href
  {http://adsabs.harvard.edu/abs/2018MNRAS.tmp..305A} {}

\bibitem[\protect\citeauthoryear{{Bandura} et~al.,}{{Bandura}
  et~al.}{2014}]{CHIMEpathfinder}
{Bandura} K.,  et~al., 2014, in Society of Photo-Optical Instrumentation
  Engineers (SPIE) Conference Series. p.~22 (\mn@eprint {arXiv} {1406.2288}),
  \mn@doi{10.1117/12.2054950}

\bibitem[\protect\citeauthoryear{{Barry}, {Hazelton}, {Sullivan}, {Morales}  \&
  {Pober}}{{Barry} et~al.}{2016}]{BarryCal}
{Barry} N.,  {Hazelton} B.,  {Sullivan} I.,  {Morales} M.~F.,   {Pober} J.~C.,
  2016, \mn@doi [\mnras] {10.1093/mnras/stw1380}, \href
  {http://adsabs.harvard.edu/abs/2016MNRAS.461.3135B} {461, 3135}

\bibitem[\protect\citeauthoryear{{Beardsley} et~al.,}{{Beardsley}
  et~al.}{2016}]{BeardsleyFirstSeason}
{Beardsley} A.~P.,  et~al., 2016, \mn@doi [\apj] {10.3847/1538-4357/833/1/102},
  \href {http://adsabs.harvard.edu/abs/2016ApJ...833..102B} {833, 102}

\bibitem[\protect\citeauthoryear{{Bernardi} et~al.,}{{Bernardi}
  et~al.}{2009}]{BernardiForegrounds}
{Bernardi} G.,  et~al., 2009, \mn@doi [\aap] {10.1051/0004-6361/200911627},
  \href {http://adsabs.harvard.edu/abs/2009A%26A...500..965B} {500, 965}

\bibitem[\protect\citeauthoryear{{Bonaldi} \& {Brown}}{{Bonaldi} \&
  {Brown}}{2015}]{BonaldiCCA}
{Bonaldi} A.,  {Brown} M.~L.,  2015, \mn@doi [\mnras] {10.1093/mnras/stu2601},
  \href {http://adsabs.harvard.edu/abs/2015MNRAS.447.1973B} {447, 1973}

\bibitem[\protect\citeauthoryear{{Chang}, {Pen}, {Peterson}  \&
  {McDonald}}{{Chang} et~al.}{2008}]{ChangDE}
{Chang} T.,  {Pen} U.,  {Peterson} J.~B.,   {McDonald} P.,  2008, \mn@doi
  [\prl.] {10.1103/PhysRevLett.100.091303}, \href
  {http://adsabs.harvard.edu/abs/2008PhRvL.100i1303C} {100, 091303}

\bibitem[\protect\citeauthoryear{{Chapman} et~al.,}{{Chapman}
  et~al.}{2013}]{ChapmanGMCA}
{Chapman} E.,  et~al., 2013, \mn@doi [\mnras] {10.1093/mnras/sts333}, \href
  {http://adsabs.harvard.edu/abs/2013MNRAS.429..165C} {429, 165}

\bibitem[\protect\citeauthoryear{{Datta}, {Bowman}  \& {Carilli}}{{Datta}
  et~al.}{2010}]{Dattapowerspec}
{Datta} A.,  {Bowman} J.~D.,   {Carilli} C.~L.,  2010, \mn@doi [ApJ]
  {10.1088/0004-637X/724/1/526}, \href
  {http://adsabs.harvard.edu/abs/2010ApJ...724..526D} {724, 526}

\bibitem[\protect\citeauthoryear{{DeBoer} et~al.,}{{DeBoer}
  et~al.}{2017}]{HERAOverview}
{DeBoer} D.~R.,  et~al., 2017, \mn@doi [\pasp]
  {10.1088/1538-3873/129/974/045001}, \href
  {http://adsabs.harvard.edu/abs/2017PASP..129d5001D} {129, 045001}

\bibitem[\protect\citeauthoryear{{Dillon} \& {Parsons}}{{Dillon} \&
  {Parsons}}{2016}]{RedArrayConfig}
{Dillon} J.~S.,  {Parsons} A.~R.,  2016, \mn@doi [\apj]
  {10.3847/0004-637X/826/2/181}, \href
  {http://adsabs.harvard.edu/abs/2016ApJ...826..181D} {826, 181}

\bibitem[\protect\citeauthoryear{{Dillon}, {Liu}  \& {Tegmark}}{{Dillon}
  et~al.}{2013}]{DillonFast}
{Dillon} J.~S.,  {Liu} A.,   {Tegmark} M.,  2013, \mn@doi [\prd]
  {10.1103/PhysRevD.87.043005}, \href
  {http://adsabs.harvard.edu/abs/2013PhRvD..87d3005D} {87, 043005}

\bibitem[\protect\citeauthoryear{{Dillon} et~al.,}{{Dillon} et~al.}{2014}]{X13}
{Dillon} J.~S.,  et~al., 2014, \mn@doi [\prd] {10.1103/PhysRevD.89.023002},
  \href {http://adsabs.harvard.edu/abs/2014PhRvD..89b3002D} {89, 023002}

\bibitem[\protect\citeauthoryear{{Dillon} et~al.,}{{Dillon}
  et~al.}{2015a}]{mapmaking}
{Dillon} J.~S.,  et~al., 2015a, \mn@doi [\prd] {10.1103/PhysRevD.91.023002},
  \href {http://adsabs.harvard.edu/abs/2015PhRvD..91b3002D} {91, 023002}

\bibitem[\protect\citeauthoryear{{Dillon} et~al.,}{{Dillon}
  et~al.}{2015b}]{EmpiricalCovariance}
{Dillon} J.~S.,  et~al., 2015b, \mn@doi [\prd] {10.1103/PhysRevD.91.123011},
  \href {http://adsabs.harvard.edu/abs/2015PhRvD..91l3011D} {91, 123011}

\bibitem[\protect\citeauthoryear{{Ewall-Wice} et~al.,}{{Ewall-Wice}
  et~al.}{2016}]{AaronFirstEoXLimits}
{Ewall-Wice} A.,  et~al., 2016, \mn@doi [\mnras] {10.1093/mnras/stw1022}, \href
  {http://adsabs.harvard.edu/abs/2016MNRAS.460.4320E} {460, 4320}

\bibitem[\protect\citeauthoryear{{Ewall-Wice}, {Dillon}, {Liu}  \&
  {Hewitt}}{{Ewall-Wice} et~al.}{2017}]{ModelingNoise}
{Ewall-Wice} A.,  {Dillon} J.~S.,  {Liu} A.,   {Hewitt} J.,  2017, \mn@doi
  [\mnras] {10.1093/mnras/stx1221}, \href
  {http://adsabs.harvard.edu/abs/2017MNRAS.470.1849E} {470, 1849}

\bibitem[\protect\citeauthoryear{{Furlanetto}, {Oh}  \& {Briggs}}{{Furlanetto}
  et~al.}{2006}]{FurlanettoReview}
{Furlanetto} S.~R.,  {Oh} S.~P.,   {Briggs} F.~H.,  2006, \mn@doi [\physrep]
  {10.1016/j.physrep.2006.08.002}, \href
  {http://adsabs.harvard.edu/abs/2006PhR...433..181F} {433, 181}

\bibitem[\protect\citeauthoryear{{Ghosh}, {Prasad}, {Bharadwaj}, {Ali}  \&
  {Chengalur}}{{Ghosh} et~al.}{2012}]{GhoshForegrounds}
{Ghosh} A.,  {Prasad} J.,  {Bharadwaj} S.,  {Ali} S.~S.,   {Chengalur} J.~N.,
  2012, \mn@doi [\mnras] {10.1111/j.1365-2966.2012.21889.x}, \href
  {http://adsabs.harvard.edu/abs/2012MNRAS.426.3295G} {426, 3295}

\bibitem[\protect\citeauthoryear{{Hamaker}, {Bregman}  \& {Sault}}{{Hamaker}
  et~al.}{1996}]{RadioPolarimetry1}
{Hamaker} J.~P.,  {Bregman} J.~D.,   {Sault} R.~J.,  1996, \aaps, \href
  {http://adsabs.harvard.edu/abs/1996A%26AS..117..137H} {117, 137}

\bibitem[\protect\citeauthoryear{{Hazelton}, {Morales}  \&
  {Sullivan}}{{Hazelton} et~al.}{2013}]{Hazelton2013}
{Hazelton} B.~J.,  {Morales} M.~F.,   {Sullivan} I.~S.,  2013, \mn@doi [\apj]
  {10.1088/0004-637X/770/2/156}, \href
  {http://adsabs.harvard.edu/abs/2013ApJ...770..156H} {770, 156}

\bibitem[\protect\citeauthoryear{{Jeli{\'c}} et~al.,}{{Jeli{\'c}}
  et~al.}{2008}]{Jelic08}
{Jeli{\'c}} V.,  et~al., 2008, \mn@doi [\mnras]
  {10.1111/j.1365-2966.2008.13634.x}, \href
  {http://adsabs.harvard.edu/abs/2008MNRAS.389.1319J} {389, 1319}

\bibitem[\protect\citeauthoryear{{Jeli{\'c}}, {Zaroubi}, {Labropoulos},
  {Bernardi}, {de Bruyn}  \& {Koopmans}}{{Jeli{\'c}}
  et~al.}{2010}]{JelicRealistic}
{Jeli{\'c}} V.,  {Zaroubi} S.,  {Labropoulos} P.,  {Bernardi} G.,  {de Bruyn}
  A.~G.,   {Koopmans} L.~V.~E.,  2010, \mn@doi [\mnras]
  {10.1111/j.1365-2966.2010.17407.x}, \href
  {http://adsabs.harvard.edu/abs/2010MNRAS.409.1647J} {409, 1647}

\bibitem[\protect\citeauthoryear{{Jones}}{{Jones}}{1941}]{Jones1941}
{Jones} R.~C.,  1941, Journal of the Optical Society of America (1917-1983),
  \href {http://adsabs.harvard.edu/abs/1941JOSA...31..488J} {31, 488}

\bibitem[\protect\citeauthoryear{{Kern}, {Carilli}, {Kohn}, {Parsons}, {Dillon}
   \& {Ali}}{{Kern} et~al.}{2017}]{kern_hera41}
{Kern} N.~S.,  {Carilli} C.~L.,  {Kohn} S.~A.,  {Parsons} A.~R.,  {Dillon}
  J.~S.,   {Ali} Z.~S.,  2017, HERA Memo \#30: Redundant Calibration
  Degeneracies with Four Polarizations

\bibitem[\protect\citeauthoryear{{Kohn} \& {Aguirre}}{{Kohn} \&
  {Aguirre}}{2015}]{kohn_hera6}
{Kohn} S.~A.,  {Aguirre} J.~E.,  2015, HERA Memo \#6: Crosstalk Measurements
  from PAPER-64

\bibitem[\protect\citeauthoryear{{Kohn} et~al.,}{{Kohn}
  et~al.}{2016}]{SaulPAPERPol}
{Kohn} S.~A.,  et~al., 2016, \mn@doi [\apj] {10.3847/0004-637X/823/2/88}, \href
  {http://adsabs.harvard.edu/abs/2016ApJ...823...88K} {823, 88}

\bibitem[\protect\citeauthoryear{{Koopmans} et~al.,}{{Koopmans}
  et~al.}{2015}]{LeonCosmicDawnEoRSKA}
{Koopmans} L.,  et~al., 2015, Advancing Astrophysics with the Square Kilometre
  Array (AASKA14), \href {http://adsabs.harvard.edu/abs/2015aska.confE...1K}
  {p.~1}

\bibitem[\protect\citeauthoryear{{Lannes} \& {Prieur}}{{Lannes} \&
  {Prieur}}{2014}]{IntegerAmbiguity}
{Lannes} A.,  {Prieur} J.-L.,  2014, \mn@doi [Astronomische Nachrichten]
  {10.1002/asna.201211947}, \href
  {http://adsabs.harvard.edu/abs/2014AN....335..198L} {335, 198}

\bibitem[\protect\citeauthoryear{{Lenc} et~al.,}{{Lenc} et~al.}{2016}]{EmilPol}
{Lenc} E.,  et~al., 2016, \mn@doi [\apj] {10.3847/0004-637X/830/1/38}, \href
  {http://adsabs.harvard.edu/abs/2016ApJ...830...38L} {830, 38}

\bibitem[\protect\citeauthoryear{{Li} et~al.,}{{Li}
  et~al.}{view}]{WenyangMWAHex}
{Li} W.,  et~al., in~review

\bibitem[\protect\citeauthoryear{{Liu} \& {Tegmark}}{{Liu} \&
  {Tegmark}}{2011}]{LT11}
{Liu} A.,  {Tegmark} M.,  2011, \mn@doi [\prd] {10.1103/PhysRevD.83.103006},
  \href {http://adsabs.harvard.edu/abs/2011PhRvD..83j3006L} {83, 103006}

\bibitem[\protect\citeauthoryear{{Liu}, {Tegmark}, {Morrison}, {Lutomirski}  \&
  {Zaldarriaga}}{{Liu} et~al.}{2010}]{redundant}
{Liu} A.,  {Tegmark} M.,  {Morrison} S.,  {Lutomirski} A.,   {Zaldarriaga} M.,
  2010, \mn@doi [\mnras] {10.1111/j.1365-2966.2010.17174.x}, \href
  {http://adsabs.harvard.edu/abs/2010MNRAS.408.1029L} {408, 1029}

\bibitem[\protect\citeauthoryear{{Liu}, {Parsons}  \& {Trott}}{{Liu}
  et~al.}{2014a}]{EoRWindow1}
{Liu} A.,  {Parsons} A.~R.,   {Trott} C.~M.,  2014a, \mn@doi [\prd]
  {10.1103/PhysRevD.90.023018}, \href
  {http://adsabs.harvard.edu/abs/2014PhRvD..90b3018L} {90, 023018}

\bibitem[\protect\citeauthoryear{{Liu}, {Parsons}  \& {Trott}}{{Liu}
  et~al.}{2014b}]{EoRWindow2}
{Liu} A.,  {Parsons} A.~R.,   {Trott} C.~M.,  2014b, \mn@doi [\prd]
  {10.1103/PhysRevD.90.023019}, \href
  {http://adsabs.harvard.edu/abs/2014PhRvD..90b3019L} {90, 023019}

\bibitem[\protect\citeauthoryear{{Loeb} \& {Furlanetto}}{{Loeb} \&
  {Furlanetto}}{2013}]{aviBook}
{Loeb} A.,  {Furlanetto} S.~R.,  2013, The First Galaxies In The Universe.
Princeton University Press, Princeton, NJ

\bibitem[\protect\citeauthoryear{{Mao}, {Tegmark}, {McQuinn}, {Zaldarriaga}  \&
  {Zahn}}{{Mao} et~al.}{2008}]{Yi}
{Mao} Y.,  {Tegmark} M.,  {McQuinn} M.,  {Zaldarriaga} M.,   {Zahn} O.,  2008,
  \mn@doi [\prd] {10.1103/PhysRevD.78.023529}, \href
  {http://adsabs.harvard.edu/abs/2008PhRvD..78b3529M} {78, 023529}

\bibitem[\protect\citeauthoryear{{Masui} et~al.,}{{Masui} et~al.}{2013}]{GBT}
{Masui} K.~W.,  et~al., 2013, \mn@doi [\apjl] {10.1088/2041-8205/763/1/L20},
  \href {http://adsabs.harvard.edu/abs/2013ApJ...763L..20M} {763, L20}

\bibitem[\protect\citeauthoryear{{McQuinn}, {Zahn}, {Zaldarriaga}, {Hernquist}
  \& {Furlanetto}}{{McQuinn} et~al.}{2006}]{Matt3}
{McQuinn} M.,  {Zahn} O.,  {Zaldarriaga} M.,  {Hernquist} L.,   {Furlanetto}
  S.~R.,  2006, \mn@doi [ApJ] {10.1086/505167}, \href
  {http://adsabs.harvard.edu/abs/2006ApJ...653..815M} {653, 815}

\bibitem[\protect\citeauthoryear{{Mishra} \& {Hirata}}{{Mishra} \&
  {Hirata}}{2017}]{21cmGravWaveCircularForecast}
{Mishra} A.,  {Hirata} C.~M.,  2017, preprint, \href
  {http://adsabs.harvard.edu/abs/2017arXiv170703514M} {} (\mn@eprint {arXiv}
  {1707.03514})

\bibitem[\protect\citeauthoryear{{Moore}, {Aguirre}, {Parsons}, {Jacobs}  \&
  {Pober}}{{Moore} et~al.}{2013}]{MoorePolarization}
{Moore} D.~F.,  {Aguirre} J.~E.,  {Parsons} A.~R.,  {Jacobs} D.~C.,   {Pober}
  J.~C.,  2013, \mn@doi [\apj] {10.1088/0004-637X/769/2/154}, \href
  {http://adsabs.harvard.edu/abs/2013ApJ...769..154M} {769, 154}

\bibitem[\protect\citeauthoryear{{Moore} et~al.,}{{Moore}
  et~al.}{2017}]{MoorePolLimits}
{Moore} D.~F.,  et~al., 2017, \mn@doi [\apj] {10.3847/1538-4357/836/2/154},
  \href {http://adsabs.harvard.edu/abs/2017ApJ...836..154M} {836, 154}

\bibitem[\protect\citeauthoryear{{Morales} \& {Wyithe}}{{Morales} \&
  {Wyithe}}{2010}]{miguelreview}
{Morales} M.~F.,  {Wyithe} J.~S.~B.,  2010, \mn@doi [\araa]
  {10.1146/annurev-astro-081309-130936}, \href
  {http://adsabs.harvard.edu/abs/2010ARA%26A..48..127M} {48, 127}

\bibitem[\protect\citeauthoryear{{Morales}, {Hazelton}, {Sullivan}  \&
  {Beardsley}}{{Morales} et~al.}{2012}]{MoralesPSShapes}
{Morales} M.~F.,  {Hazelton} B.,  {Sullivan} I.,   {Beardsley} A.,  2012,
  \mn@doi [\apj] {10.1088/0004-637X/752/2/137}, \href
  {http://adsabs.harvard.edu/abs/2012ApJ...752..137M} {752, 137}

\bibitem[\protect\citeauthoryear{{Newburgh} et~al.,}{{Newburgh}
  et~al.}{2016}]{HIRAXconcept}
{Newburgh} L.~B.,  et~al., 2016, in Ground-based and Airborne Telescopes VI. p.
  99065X (\mn@eprint {arXiv} {1607.02059}), \mn@doi{10.1117/12.2234286}

\bibitem[\protect\citeauthoryear{{Noorishad}, {Wijnholds}, {van Ardenne}  \&
  {van der Hulst}}{{Noorishad} et~al.}{2012}]{LOFARcal}
{Noorishad} P.,  {Wijnholds} S.~J.,  {van Ardenne} A.,   {van der Hulst} J.~M.,
   2012, \mn@doi [\aap] {10.1051/0004-6361/201219087}, \href
  {http://adsabs.harvard.edu/abs/2012A%26A...545A.108N} {545, A108}

\bibitem[\protect\citeauthoryear{{Nunhokee} et~al.,}{{Nunhokee}
  et~al.}{2017}]{NunhokeePolarized}
{Nunhokee} C.~D.,  et~al., 2017, \mn@doi [\apj] {10.3847/1538-4357/aa8b73},
  \href {http://adsabs.harvard.edu/abs/2017ApJ...848...47N} {848, 47}

\bibitem[\protect\citeauthoryear{{Paciga} et~al.,}{{Paciga}
  et~al.}{2013}]{newGMRT}
{Paciga} G.,  et~al., 2013, \mn@doi [\mnras] {10.1093/mnras/stt753}, \href
  {http://adsabs.harvard.edu/abs/2013MNRAS.433..639P} {433, 639}

\bibitem[\protect\citeauthoryear{{Parsons}, {Pober}, {McQuinn}, {Jacobs}  \&
  {Aguirre}}{{Parsons} et~al.}{2012a}]{AaronSensitivity}
{Parsons} A.,  {Pober} J.,  {McQuinn} M.,  {Jacobs} D.,   {Aguirre} J.,  2012a,
  \mn@doi [\apj] {10.1088/0004-637X/753/1/81}, \href
  {http://adsabs.harvard.edu/abs/2012ApJ...753...81P} {753, 81}

\bibitem[\protect\citeauthoryear{{Parsons}, {Pober}, {Aguirre}, {Carilli},
  {Jacobs}  \& {Moore}}{{Parsons} et~al.}{2012b}]{AaronDelay}
{Parsons} A.~R.,  {Pober} J.~C.,  {Aguirre} J.~E.,  {Carilli} C.~L.,  {Jacobs}
  D.~C.,   {Moore} D.~F.,  2012b, \mn@doi [\apj] {10.1088/0004-637X/756/2/165},
  \href {http://adsabs.harvard.edu/abs/2012ApJ...756..165P} {756, 165}

\bibitem[\protect\citeauthoryear{{Parsons} et~al.,}{{Parsons}
  et~al.}{2014}]{PAPER32Limits}
{Parsons} A.~R.,  et~al., 2014, \mn@doi [\apj] {10.1088/0004-637X/788/2/106},
  \href {http://adsabs.harvard.edu/abs/2014ApJ...788..106P} {788, 106}

\bibitem[\protect\citeauthoryear{{Parsons}, {Liu}, {Ali}  \& {Cheng}}{{Parsons}
  et~al.}{2016}]{FringeRateFilter}
{Parsons} A.~R.,  {Liu} A.,  {Ali} Z.~S.,   {Cheng} C.,  2016, \mn@doi [\apj]
  {10.3847/0004-637X/820/1/51}, \href
  {http://adsabs.harvard.edu/abs/2016ApJ...820...51P} {820, 51}

\bibitem[\protect\citeauthoryear{{Patil} et~al.,}{{Patil}
  et~al.}{2017}]{LOFARLimitsPatil}
{Patil} A.~H.,  et~al., 2017, \mn@doi [\apj] {10.3847/1538-4357/aa63e7}, \href
  {http://adsabs.harvard.edu/abs/2017ApJ...838...65P} {838, 65}

\bibitem[\protect\citeauthoryear{{Pearson} \& {Readhead}}{{Pearson} \&
  {Readhead}}{1984}]{selfcal}
{Pearson} T.~J.,  {Readhead} A.~C.~S.,  1984, \mn@doi [\araa]
  {10.1146/annurev.aa.22.090184.000525}, \href
  {http://adsabs.harvard.edu/abs/1984ARA%26A..22...97P} {22, 97}

\bibitem[\protect\citeauthoryear{{Pober} et~al.,}{{Pober}
  et~al.}{2013}]{PoberWedge}
{Pober} J.~C.,  et~al., 2013, \mn@doi [\apjl] {10.1088/2041-8205/768/2/L36},
  \href {http://adsabs.harvard.edu/abs/2013ApJ...768L..36P} {768, L36}

\bibitem[\protect\citeauthoryear{{Pober} et~al.,}{{Pober}
  et~al.}{2014}]{PoberNextGen}
{Pober} J.~C.,  et~al., 2014, \mn@doi [\apj] {10.1088/0004-637X/782/2/66},
  \href {http://adsabs.harvard.edu/abs/2014ApJ...782...66P} {782, 66}

\bibitem[\protect\citeauthoryear{{Pritchard} \& {Loeb}}{{Pritchard} \&
  {Loeb}}{2012}]{PritchardLoebReview}
{Pritchard} J.~R.,  {Loeb} A.,  2012, \mn@doi [Reports on Progress in Physics]
  {10.1088/0034-4885/75/8/086901}, \href
  {http://adsabs.harvard.edu/abs/2012RPPh...75h6901P} {75, 086901}

\bibitem[\protect\citeauthoryear{{Ram Marthi} \& {Chengalur}}{{Ram Marthi} \&
  {Chengalur}}{2013}]{NonLinearRedCal}
{Ram Marthi} V.,  {Chengalur} J.,  2013, preprint, \href
  {http://adsabs.harvard.edu/abs/2013arXiv1310.1449R} {} (\mn@eprint {arXiv}
  {1310.1449})

\bibitem[\protect\citeauthoryear{{Santos}, {Cooray}  \& {Knox}}{{Santos}
  et~al.}{2005}]{Santos}
{Santos} M.~G.,  {Cooray} A.,   {Knox} L.,  2005, \mn@doi [ApJ]
  {10.1086/429857}, \href {http://adsabs.harvard.edu/abs/2005ApJ...625..575S}
  {625, 575}

\bibitem[\protect\citeauthoryear{{Sault}, {Hamaker}  \& {Bregman}}{{Sault}
  et~al.}{1996}]{RadioPolarimetry2}
{Sault} R.~J.,  {Hamaker} J.~P.,   {Bregman} J.~D.,  1996, \aaps, \href
  {http://adsabs.harvard.edu/abs/1996A%26AS..117..149S} {117, 149}

\bibitem[\protect\citeauthoryear{{Shaw}, {Sigurdson}, {Pen}, {Stebbins}  \&
  {Sitwell}}{{Shaw} et~al.}{2014}]{Richard}
{Shaw} J.~R.,  {Sigurdson} K.,  {Pen} U.-L.,  {Stebbins} A.,   {Sitwell} M.,
  2014, \mn@doi [\apj] {10.1088/0004-637X/781/2/57}, \href
  {http://adsabs.harvard.edu/abs/2014ApJ...781...57S} {781, 57}

\bibitem[\protect\citeauthoryear{{Shaw}, {Sigurdson}, {Sitwell}, {Stebbins}  \&
  {Pen}}{{Shaw} et~al.}{2015}]{ShawCoaxing}
{Shaw} J.~R.,  {Sigurdson} K.,  {Sitwell} M.,  {Stebbins} A.,   {Pen} U.-L.,
  2015, \mn@doi [\prd] {10.1103/PhysRevD.91.083514}, \href
  {http://adsabs.harvard.edu/abs/2015PhRvD..91h3514S} {91, 083514}

\bibitem[\protect\citeauthoryear{{Sievers}}{{Sievers}}{2017}]{CorrCal}
{Sievers} J.~L.,  2017, preprint, \href
  {http://adsabs.harvard.edu/abs/2017arXiv170101860S} {} (\mn@eprint {arXiv}
  {1701.01860})

\bibitem[\protect\citeauthoryear{{Thompson}, {Moran}  \& {Swenson}}{{Thompson}
  et~al.}{2017}]{ThompsonMoranSwenson}
{Thompson} A.~R.,  {Moran} J.~M.,   {Swenson} Jr. G.~W.,  2017, {Interferometry
  and Synthesis in Radio Astronomy, 3rd Edition},
  \mn@doi{10.1007/978-3-319-44431-4.
}

\bibitem[\protect\citeauthoryear{{Thyagarajan} et~al.,}{{Thyagarajan}
  et~al.}{2013}]{ThyagarajanWedge}
{Thyagarajan} N.,  et~al., 2013, \mn@doi [\apj] {10.1088/0004-637X/776/1/6},
  \href {http://adsabs.harvard.edu/abs/2013ApJ...776....6T} {776, 6}

\bibitem[\protect\citeauthoryear{{Trott} et~al.,}{{Trott} et~al.}{2016}]{CHIPS}
{Trott} C.~M.,  et~al., 2016, \mn@doi [\apj] {10.3847/0004-637X/818/2/139},
  \href {http://adsabs.harvard.edu/abs/2016ApJ...818..139T} {818, 139}

\bibitem[\protect\citeauthoryear{{Vedantham} \& {Koopmans}}{{Vedantham} \&
  {Koopmans}}{2015}]{HarishLeonIon1}
{Vedantham} H.~K.,  {Koopmans} L.~V.~E.,  2015, \mn@doi [\mnras]
  {10.1093/mnras/stv1594}, \href
  {http://adsabs.harvard.edu/abs/2015MNRAS.453..925V} {453, 925}

\bibitem[\protect\citeauthoryear{{Vedantham}, {Udaya Shankar}  \&
  {Subrahmanyan}}{{Vedantham} et~al.}{2012}]{VedanthamWedge}
{Vedantham} H.,  {Udaya Shankar} N.,   {Subrahmanyan} R.,  2012, \mn@doi [\apj]
  {10.1088/0004-637X/745/2/176}, \href
  {http://adsabs.harvard.edu/abs/2012ApJ...745..176V} {745, 176}

\bibitem[\protect\citeauthoryear{{Wieringa}}{{Wieringa}}{1992}]{Wieringa}
{Wieringa} M.~H.,  1992, \mn@doi [Experimental Astronomy] {10.1007/BF00420576},
  \href {http://adsabs.harvard.edu/abs/1992ExA.....2..203W} {2, 203}

\bibitem[\protect\citeauthoryear{{Wyithe}, {Loeb}  \& {Geil}}{{Wyithe}
  et~al.}{2008}]{wyithe2008}
{Wyithe} J.~S.~B.,  {Loeb} A.,   {Geil} P.~M.,  2008, \mn@doi [MNRAS]
  {10.1111/j.1365-2966.2007.12631.x}, \href
  {http://adsabs.harvard.edu/abs/2008MNRAS.383.1195W} {383, 1195}

\bibitem[\protect\citeauthoryear{{Yatawatta}}{{Yatawatta}}{2012}]{YatawattaAmbiguity}
{Yatawatta} S.,  2012, \mn@doi [Experimental Astronomy]
  {10.1007/s10686-012-9300-7}, \href
  {http://adsabs.harvard.edu/abs/2012ExA....34...89Y} {34, 89}

\bibitem[\protect\citeauthoryear{{Yatawatta}}{{Yatawatta}}{2016}]{YatawattaConsensus}
{Yatawatta} S.,  2016, preprint, \href
  {http://adsabs.harvard.edu/abs/2016arXiv160509219Y} {} (\mn@eprint {arXiv}
  {1605.09219})

\bibitem[\protect\citeauthoryear{{Yatawatta} et~al.,}{{Yatawatta}
  et~al.}{2013}]{InitialLOFAR1}
{Yatawatta} S.,  et~al., 2013, \mn@doi [\aap] {10.1051/0004-6361/201220874},
  \href {http://adsabs.harvard.edu/abs/2013A%26A...550A.136Y} {550, A136}

\bibitem[\protect\citeauthoryear{{Zaroubi}}{{Zaroubi}}{2013}]{SaleemEoRChapter}
{Zaroubi} S.,  2013, in {Wiklind} T.,  {Mobasher} B.,   {Bromm} V.,  eds,
  Astrophysics and Space Science Library Vol. 396, The First Galaxies. p.~45
  (\mn@eprint {arXiv} {1206.0267}), \mn@doi{10.1007/978-3-642-32362-1_2}

\bibitem[\protect\citeauthoryear{{Zheng} et~al.,}{{Zheng}
  et~al.}{2014}]{MITEoR}
{Zheng} H.,  et~al., 2014, \mn@doi [\mnras] {10.1093/mnras/stu1773}, \href
  {http://adsabs.harvard.edu/abs/2014MNRAS.445.1084Z} {445, 1084}

\bibitem[\protect\citeauthoryear{{Zheng} et~al.,}{{Zheng}
  et~al.}{2017}]{ZhengBruteForce}
{Zheng} H.,  et~al., 2017, \mn@doi [\mnras] {10.1093/mnras/stw2910}, \href
  {http://adsabs.harvard.edu/abs/2017MNRAS.465.2901Z} {465, 2901}

\makeatother
\end{thebibliography}

\label{lastpage}
\end{document}